\documentclass[aps,pra,floatfix,showpacs,twocolumn,tightenlines,superscriptaddress]{revtex4}

\usepackage{amsmath}
\usepackage[hypertex]{hyperref}
\usepackage{graphicx}
\usepackage{bm}
\usepackage{amsfonts}
\usepackage{amsthm}

\newtheorem{theorem}{Theorem}[section]

\newtheorem{lemma}{Lemma}[section]
\newtheorem{corollary}{Corollary}[theorem]
\newtheorem*{problem}{Problem}

\def\bra#1{\mathinner{\langle{#1}|}}
\def\ket#1{\mathinner{|{#1}\rangle}}

\begin{document}

\title{Optimal tracking for pairs of qubit states}

\author{Paulo E. M. F.~Mendon{\c c}a}\email{mendonca@physics.uq.edu.au}
\affiliation{Department of Physics, The University of Queensland, Queensland 4072, Australia}

\author{Alexei Gilchrist}\email{alexei@ics.mq.edu.au}
\affiliation{Physics Department, Macquarie University, Sydney, NSW 2109, Australia}

\author{Andrew C. Doherty}\email{doherty@physics.uq.edu.au}
\affiliation{Department of Physics, The University of Queensland, Queensland 4072, Australia}
\date{\today}

\begin{abstract}
In classical control theory, tracking refers to the ability to perform measurements and feedback on a classical system in order to enforce some desired dynamics. In this paper we investigate a simple version of quantum tracking, namely, we look at how to optimally transform the state of a single qubit into a given target state, when the system can be prepared in two different ways, and the target state depends on the choice of preparation. We propose a tracking strategy that is proved to be optimal for any input and target states. Applications in the context of state discrimination, state purification, state stabilization and state-dependent quantum cloning are presented, where existing optimality results are recovered and extended.
\end{abstract}

\pacs{03.67.Pp, 03.65.Ta, 03.67.-a}

\maketitle

\section{Introduction}
\label{sec:intro}

A common goal of many problems in quantum information science is the search for quantum operations that simultaneously transform a set of given input quantum states into another pre-specified set. Well known examples are task such as quantum cloning, state discrimination and quantum error correction.\par

In general, though, quantum mechanics forbids arbitrary quantum state dynamics. As a result, one is left with several examples of ``impossible quantum machines'' \cite{01Werner}. Not only is quantum cloning unachievable \cite{82Dieks271,82Wootters802,96Barnum2818}, but also quantum state discrimination strategies are typically subject to non-zero misidentification probabilities \cite{76Helstrom} and/or inconclusive outcomes \cite{88Dieks303,87Ivanovic257,88Peres19} and there are no quantum error correction protocols capable of fully reverting the action of an arbitrary noise model \cite{03Gregoratti915}.\par

Nevertheless, it is still possible to approximate \emph{ideal} (but unphysical) transformations with \emph{optimal} (but physical) ones. This provides quantum limits to the performance of tasks such as state discrimination, cloning and so on. In this paper, we study the general problem of transforming the state of a single qubit into a given target state, when the system can be prepared in two different ways, and the target state depends on the choice of preparation. We call this task \emph{quantum tracking}, a term borrowed from classical control theory. Our main result is an analytical description of an optimal quantum tracking strategy.\par

More specifically, the quantum tracking problem studied here can be understood as follows. Consider that Alice prepares either a qubit state $\rho_1$ with probability $\pi_1$ or $\rho_2$ with probability $\pi_2$. Bob is allowed to interact with the system in any physically allowed way, aiming to enforce the tracking rule
\begin{equation}\label{rule}
\mbox{if Alice prepared } \rho_i\,, \mbox{ then ouput } \overline{\rho}_i
\end{equation}
for $i=1,2$ and some given qubit density matrices $\overline{\rho}_i$.\par

At his disposal, Bob has all the information about the possible preparations $\rho_i$ and their respective prior probabilities $\pi_i$, but \emph{not} the actual preparation (the value of the index $i$).

Because quantum states are generally not perfectly distinguishable, a strategy that attempts to identify Alice's preparation and then reprepare the target according to rule (\ref{rule}) is not always guaranteed to succeed. In fact,
 this limited distinguishability is an unsurpassable obstacle in the implementation of (\ref{rule}).

In this paper, an optimal solution is defined as follows.
Amongst all the physical transformations acting on the input states $\rho_i$, an optimal one is any map that outputs density matrices $\rho_i^\prime$ such that the averaged Hilbert-Schmidt inner product between $\rho_i^\prime$ and $\overline{\rho}_i$ is maximal. When $\overline{\rho}_i$ are pure states, such a figure-of-merit coincides with the averaged Uhlmann-Jozsa fidelity \cite{94Jozsa2315,76Uhlmann273}, and this notion of optimality gains an appealing operational interpretation \cite{95Schumacher2738}. Suppose that Alice [aware of her preparation and of rule (\ref{rule})], decides to check whether Bob prepared the density matrix he was supposed to, and for that purpose she performs a verification measurement on the density matrix produced by him. If Bob chooses an optimal transformation according to the above prescription, then the probability he will pass Alice's test is as large as allowed by quantum mechanics.

The tracking problem resembles the transformability problem for pairs of qubit states studied by Alberti and Uhlmann in the 80's \cite{80Alberti163} (see also Appendix \ref{app:ptc}). In \cite{80Alberti163}, a criterion based on the distinguishability between the source density matrices and the distinguishability between the target density matrices was developed in order to decide on the existence of a completely positive and trace preserving (CPTP) map simultaneously transforming each source into each target.\par

\begin{figure}
\includegraphics[width=8.5cm]{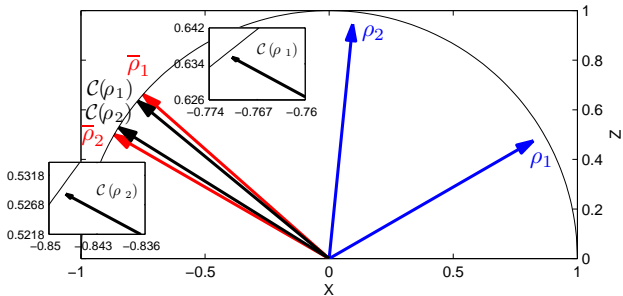} 
\caption{(Color online) The transformation of the mixed states $\rho_1$ and $\rho_2$ (in blue) into the
pure states $\overline{\rho}_1$ and $\overline{\rho}_2$ (in red) is not a physical one. However, there is a physical transformation $\mathcal{C}$ capable of transforming the input states into states $\mathcal{C}(\rho_i)$ (in black) which closely approximates the targets. Note, in the detail, that $\mathcal{C}(\rho_1)$ and $\mathcal{C}(\rho_2)$ are still slightly mixed.}\label{fig:approx}
\end{figure}

Although Alberti and Uhlmann's criterion classifies the set of states $\rho_1$, $\rho_2$, $\overline{\rho}_1$ and $\overline{\rho}_2$ for which rule (\ref{rule}) can be satisfied, it does not provide a construction of the CPTP map implementing that transformation, nor touch the problem of how to find a feasible approximation when the criterion is not satisfied. For many purposes, the requirement of perfectly converting sources into targets is unnecessarily strong, as some strictly impossible physical transformations can be very well approximated by physical ones, as illustrated in Fig \ref{fig:approx}. In fact, any experimental realization of a map is just an approximation of it.

 \par

Another problem closely related to the aims of this paper was investigated in Ref. \cite{07Branczyk012329}. Specifically, we considered the problem of determining the optimal quantum operation to stabilize the state of a single qubit, randomly prepared in one of two pure states, against the effect of dephasing noise. The results of \cite{07Branczyk012329} are here extended in several ways. The input states are allowed to be mixed and prepared with arbitrary prior probability distribution; the noise model is arbitrary and, most importantly, the stabilization task is replaced with tracking.\par

Finally, there is an intrinsic connection between the quantum tracking problem and the ``optimization approach'' \cite{05Reimpell080501,07Fletcher012338, 06Reimpell, 06Kosut,08Kosut020502,07Yamamoto012327,05Yamamoto022322} to quantum error correction \footnote{This is in contrast with the traditional approach to quantum error correction, which followed the direction of adapting classical coding techniques to the quantum domain \cite{97Knill900}.}. In these references, the encoding and recovery operations are regarded as optimization variables whose optimal values maximize a given figure-of-merit (typically a function bounded between $0$ and $1$, equal to $1$ if and only if the noise dynamics is reversible). Efficient numerical methods are then proposed to solve the optimization problem. The key differences between our work and these references is that we do not consider encoding of the initial state and focus on reverting the noise dynamics experienced by only a pair of states. By doing so, the optimization of the recovery operation can be handled analytically for a conveniently chosen figure-of-merit.\par

The paper is structured as follows. Section \ref{sec:probandstrategy} introduces the formal statement of the problem and our working strategy, which is proved to be optimal in Section \ref{sec:optimal}. In section \ref{sec:examples} we evaluate the performance of the optimal strategy in the contexts of quantum state discrimination, quantum state stabilization in the presence noise, perfect quantum tracking, and state-dependent quantum cloning. Section \ref{sec:quantctrl} proposes a physical implementation of our strategy in terms of closed and open loop control. Section \ref{sec:discussion} discusses generalizations of the problem and concludes the paper.

\section{Problem and Strategy}\label{sec:probandstrategy}
In this section we give a formal statement of the problem of interest and introduce our strategy
\subsection{The Problem}\label{sec:problem}
Formally, the problem we set out to solve can be stated as
\begin{problem}\label{problem}
Given qubit density matrices $\rho_1$, $\rho_2$, $\overline{\rho}_1$, $\overline{\rho}_2$ (with $\rho_1\neq\rho_2$) and probabilities $\pi_1$, $\pi_2$ with $\pi_1+\pi_2=1$, find a quantum operation $\mathcal{C}$ maximizing
\begin{equation}\label{eq:J}
\mathcal{F}_{\rm HS}=\pi_1{\rm Tr}\left[\mathcal{C}(\rho_1)\overline{\rho}_1\right]+\pi_2{\rm Tr}\left[\mathcal{C}(\rho_2)\overline{\rho}_2\right]\,.
\end{equation}
\end{problem}
\noindent We will refer to this as ``the tracking problem''.\par

The choice of the average Hilbert-Schmidt inner product as our figure-of-merit $\mathcal{F}_{\rm HS}$ is motivated by technical reasons (to be clarified later), and by the fact that for pure target states (the case of greater interest as far as applications are concerned), $\mathcal{F}_{\rm HS}$ is precisely equal to the average fidelity. When the target states are mixed, $\mathcal{F}_{\rm HS}$ is a lower bound to the average fidelity \cite{94Jozsa2315}. Although not as well motivated as in the case of pure target states, the determination of the quantum operation $\mathcal{C}$ maximizing $\mathcal{F}_{\rm HS}$ can still be useful for mixed target states. For example, if a certain application requires tracking to be performed with average fidelity $f$ and the optimal value of our figure-of-merit is such that $\mathcal{F}_{\rm HS}\geq f$, then $\mathcal{C}$ is suitable for the task.\par

As a final remark, note that we do not exclude the case $\overline{\rho}_1=\overline{\rho}_2$ from the statement of the problem. However, we will exclude the case $\overline{\rho}_1=\overline{\rho}_2 = \openone/2$ from the following analysis. Obviously, this particular transformation is always feasible and achieved with the completely depolarizing channel.\par

Next, we propose a strategy that will later be proved to be a solution of this tracking problem.

\subsection{The Strategy}\label{sec:strategy}

In this section, we provide an analytical solution of the tracking problem, i.e., we detail the structure of an optimal tracking operation $\mathcal{C}$ and derive closed forms for the associated maximal value of the figure-of-merit $\mathcal{F}_{\rm HS}$. The scheme proposed here was constructed by incorporating some features observed from the numerical solution of the tracking problem in an analytical optimization procedure. In the next section, we will show that the tracking problem can be cast as a semidefinite program (SDP) \cite{96Vandenberghe49,04boyd}, and will employ the theory for this type of optimization problem to prove that the strategy presented here actually solves the tracking problem.

A quantum operation is a description of a certain physically allowed evolution of a quantum state. For a closed quantum system (not interacting with an environment) this description is given by the familiar unitary evolution of Schr\"odinger's equation. For open quantum systems, unitary evolution alone does not account for every possible state transformation --- in this case, the set of quantum operations is identified with the more comprehensive set of completely positive and trace preserving (CPTP) maps.

Any one qubit CPTP map $\mathcal{C}$ can be decomposed as \cite{01King192,02Ruskai159}
\begin{equation}\label{eq:ruskai_decomposition} \mathcal{C}(\varsigma)=U
\mathcal{D} (V \varsigma V^\dagger) U^\dagger\,,
\end{equation} where $U$, $V$ are
unitary matrices and $\mathcal{D}$ induces an affine transformation on the
input Bloch vectors; namely, it contracts the $x$, $y$ and $z$ components via a
multiplicative factor, and subsequently adds a fixed number to them. Any
non-unitary evolution arises from a transformation of this type. In the
framework of Eq. (\ref{eq:ruskai_decomposition}), unitary dynamics is simply
obtained by making the affine map $\mathcal{D}$ redundant (e.g., multiplying by
1's and adding 0's to the $x-$, $y-$ and $z-$Bloch components).

In general, CPTP maps reduce the distinguishability of quantum states. On the
Bloch sphere this typically corresponds to a reduction of the Bloch vector
length and angles between vectors. In contrast unitary dynamics preserves
the angles between Bloch vectors and their lengths.
For the tracking problem, we can imagine that in
some cases the optimal strategy will preserve lengths and angles, i.e. it will
be some unitary correction. We will construct an ``indicator function'' which
will flag this case.

\subsubsection{Indicator function}\label{sec:indicator}
To gain some intuition, we start by constructing an indicator function for the simplest case of tracking with uniform priorities $\pi_1=\pi_2=1/2$ from pure states ($\rho_1$ and $\rho_2$) to pure states ($\overline{\rho}_1$ and $\overline{\rho}_2$). Throughout, $\Theta$ and $\overline{\Theta}$ will denote the angles between the Bloch vectors of $\rho_1$, $\rho_2$ and $\overline{\rho}_1$, $\overline{\rho}_2$, respectively. It will also be convenient to define $\theta$ and $\overline{\theta}$ to be the half-angle between the Bloch vectors, i.e., $2\theta=\Theta$ and $2\overline{\theta}=\overline{\Theta}$.\par

Given that all the states involved are pure, it is straightforward to conclude that if $\Theta=\overline{\Theta}$, then unitary dynamics is the best choice --- a suitable rotation of the Bloch vectors of the inputs can perfectly bring them to coincide with the Bloch vectors of the targets (as opposed to a non-unitary evolution that would decrease the angle, hence excluding the possibility of perfect tracking).\par

 A corollary of a theorem by Alberti and Uhlmann \cite{80Alberti163} (see appendix \ref{app:ptc}) implies that any pure state transformation such that $\Theta>\overline{\Theta}$, can be perfectly implemented. If that is the case, then this transformation must be non-unitary, since a unitary would not be able to bring the angles to perfectly match. This suggests the introduction of the function
\begin{equation}\label{eq:IF_simple}
\widetilde{\Omega}\mathrel{\mathop:}=\Theta-\overline{\Theta}\,,
\end{equation}
to indicate non-unitary dynamics whenever $\widetilde{\Omega}>0$. Next, we argue that $\widetilde{\Omega}\leq 0$ indicates unitary dynamics, thus establishing $\widetilde{\Omega}$ as an example of indicator function we were looking for.\par

 We have already seen that $\widetilde{\Omega}=0$ implies unitary dynamics. Intuitively, this conclusion can be extended to $\widetilde{\Omega}<0$ with the following reasoning. If $\Theta<\overline{\Theta}$, any further decrease of the initial angle can only further separate the resulting states from the targets. Since there is not any quantum operation capable of increasing this angle, the best policy must be to preserve it, hence a unitary.

The above discussion may suggest that the optimal indicator function is merely a comparison of the
distinguishabilities between sources and targets. If the sources are more distinguishable than the targets, then we employ a quantum measurement to decrease the distinguishability, hence approximating the targets. If the sources are no more distinguishable than the targets, then we employ a unitary operation to avoid a further decrease of the overlap between the output states and the targets.
Although this reasoning is certainly in agreement with the indicator function introduced above for the special case of
pure states, it does not extend to mixed state transformations \footnote{At least not as far as the optimization of the figure-of-merit of Eq. (\ref{eq:J}) is concerned. A possibly interesting problem would be the determination of a figure-of-merit that would preserve such behavior for general state transformations.}.

If the states are not pure and the priorities are not uniform, it is much more difficult to understand how purities, angles and priorities combine to form a meaningful decision criterion about the nature of the best dynamics. In order to introduce an indicator function for this general case (obtained from some mathematical optimization procedure, not from an heuristic argument), we first define some useful notation.\par

Let $\bm{R}_i$ and $\overline{\bm{R}}_i$ be the Bloch vectors for $\rho_i$ and $\pi_i\overline{\rho}_i$, respectively (note that $\overline{\bm{R}}_i$ is not normalized). Symbolically, for $\bm{\mathcal{R}}\in\{\bm{R},\overline{\bm{R}}\}$, define
\begin{subequations}
\begin{align}
\bm{\mathcal{R}}_+&\mathrel{\mathop:}=\bm{\mathcal{R}}_1+\bm{\mathcal{R}}_2\,,\\
\bm{\mathcal{R}}_-&\mathrel{\mathop:}=\bm{\mathcal{R}}_1-\bm{\mathcal{R}}_2\,,\\
\bm{\mathcal{R}}_\times&\mathrel{\mathop:}=\bm{\mathcal{R}}_1\times\bm{\mathcal{R}}_2\,,\label{eq:Rx}
\end{align}
\end{subequations}
and as usual, the corresponding unbolded type gives the Euclidean norm $\mathcal{R}_{\{+,-,\times\}}=\|\bm{\mathcal{R}}_{\{+,-,\times\}}\|$. Also, the following will be important throughout
\begin{align}
T&\mathrel{\mathop:}= \sum_{i,j=1}^{2}{(1-\bm{R}_i\cdot \bm{R}_j)(\overline{\bm{R}}_i\cdot\overline{\bm{R}}_j)}\,,\label{eq:T}\\
S&\mathrel{\mathop:}= \sqrt{T^2+4\overline{R}_\times^2(R_-^2-R_\times^2)}\,.\label{eq:S}
\end{align}

We note that $\bm{R}_1\neq\bm{R}_2$ guarantees that $R_-^2-R_\times^2 > 0$ (see Appendix \ref{app:ST}, Lemma \ref{lemma1}), hence both $S$ and $T$ are real numbers.

In terms of these quantities, we define
 \begin{equation}\label{eq:IF}
 \Omega\mathrel{\mathop:}= S+T-2\overline{R}_\times R_\times\,,
 \end{equation}
with $\Omega>0$ indicating that non-unitary dynamics are required (which will be detailed as ``procedure A'') and $\Omega\leq 0$ indicating that unitary dynamics (``procedure B'') are required. \par

\begin{figure}
\begin{center}
\includegraphics[width=8cm]{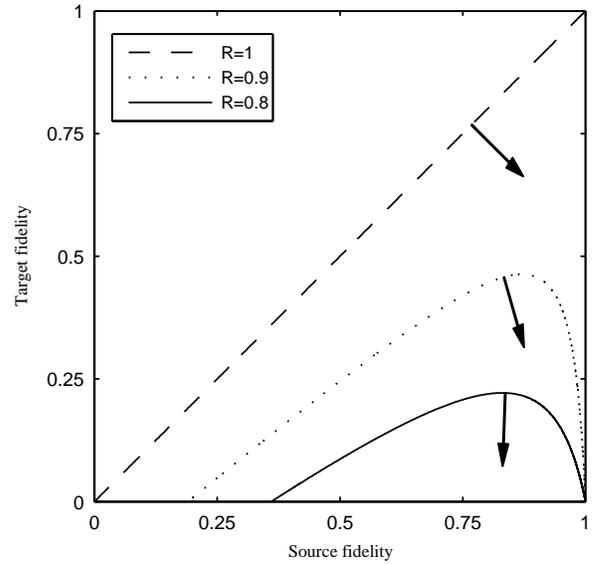}
\end{center}
\caption{Lines separating unitary and non-unitary dynamics, as prescribed by the indicator function of Eq. (\ref{eq:IF}).
For source states, we consider pairs of mixed states with Bloch vector length $R$ and separated by an angle $\Theta=2\theta$ (Bloch sphere angle).
For target states, we consider pairs of pure states separated by an angle $\overline{\Theta}=2\overline{\theta}$. The fidelity between the source states (horizontal axis) is $1-R^2\sin^2{\theta}$, and the fidelity between the target states (vertical axis) is $\cos^2{\overline{\theta}}$. The region where unitary dynamics is advisable ($\Omega\leq 0$) is indicated with an arrow. The intuitive notion that a measurement is employed when the targets are less distinguishable than the sources (and a unitary, otherwise), only holds if $R=1$ (pure sources). For $R=0.9, 0.8$ the division line moves down in such a way as to increase the portion of the parameter space where non-unitary dynamics is advisable $(\Omega > 0)$.}\label{fig:omegavsdist}
\end{figure}

 Although it would be difficult to motivate the indicator function $\Omega$ of Eq. (\ref{eq:IF}) as we did with $\widetilde{\Omega}$ in Eq. (\ref{eq:IF_simple}), it is possible to see that the former is equivalent to the latter in the case of pure qubit states. This is shown in Fig. \ref{fig:omegavsdist}, where it is also noticeable that even a simple generalization of the input states from pure to mixed states with the same level of mixedness (as measured by the norm of their Bloch vector $R$), is already sufficient to give a fairly non-trivial division line between the two types of dynamics.

\subsubsection{Procedure A}
In this section we present the details of the map $\mathcal{C}$ from Eq. (\ref{eq:ruskai_decomposition}) for $\Omega>0$ (which indicates non-unitary dynamics).

\paragraph*{Step 1.} The rotation by the unitary $V$ takes the two input Bloch vectors to vectors $\bm{R}_1^\prime$ and $\bm{R}_2^\prime$ in the $xz$-plane in such a way that they share a common positive $\bm{x}$-component and $\bm{R}_1^\prime\cdot\bm{z}>\bm{R}_2^\prime\cdot\bm{z}$, explicitly
\begin{equation}\label{eq:V}
\bm{R}_i\stackrel{V}{\longmapsto}\bm{R}_i^\prime=\frac{R_\times}{R_-}\bm{x}+\frac{\left(\bm{R}_i\cdot\bm{R}_-\right)}{R_-}\bm{z}\,.
\end{equation}

\paragraph*{Step 2.} The affine transformation $\mathcal{D}$ shortens the $\bm{x}$, $\bm{y}$ and $\bm{z}$ components of its inputs by multiplying them, respectively, by $\mu_1$, $\mu_2$, $\mu_3$ with $0\leq\mu_{\{1,2,3\}}\leq 1$ and subsequently adding $s_1$ to the $\bm{x}$ component. Applied to $\bm{R}_i^\prime$, that reads
\begin{equation}\label{eq:rill}
\bm{R}_i^\prime\stackrel{\mathcal{D}}{\longmapsto}\bm{R}_i^{\prime\prime}= \left(s_1+\mu_1\frac{R_\times}{R_-}\right)\bm{x}+\mu_3\frac{\left(\bm{R}_i\cdot\bm{R}_-\right)}{R_-}\bm{z}\,.
\end{equation}
That such a transformation can be physically implemented is not a trivial fact. Indeed, strict conditions involving the parameters $\mu_{\{1,2,3\}}$ and $s_1$ must be satisfied to guarantee the feasibility of transformation (\ref{eq:rill}) as a CPTP map \cite{02Ruskai159}. The following values can be shown to satisfy these conditions

\begin{subequations}\label{eq:nontrivial}
\begin{align}
\mu_1&=2\sqrt{\frac{2}{S(S+T)^{3}}}\overline{R}_\times^2 R_\times R_-\,,\\
\mu_2&=\left(\frac{2}{S+T}\right)\overline{R}_\times R_\times\,,\\
\mu_3&=\sqrt{\frac{2}{S(S+T)}}\overline{R}_\times R_-\,,\\
s_1&=\sqrt{\frac{1}{2S(S+T)^{3}}}\left[(S+T)^2-4\overline{R}_\times^2 R_\times^2\right]\,.
\end{align}
\end{subequations}

In Appendix \ref{app:ST} we show that the only circumstances under which the inequalities $S>0$ and $S+T>0$ are not simultaneously satisfied have $\Omega=0$. Therefore, the quantities above are real and well-defined for the present procedure ($\Omega>0$). It is not difficult to check that $\mu_1=\mu_2\mu_3$ and $s_1=\sqrt{(1-\mu_2^2)(1-\mu_3^2)}$, so the resulting map acting on density matrices is an extremal point of the convex set of CPTP maps \cite{02Ruskai159}.\par

Remarkably, if the target Bloch vectors $\overline{\bm{R}}_1$ and $\overline{\bm{R}}_2$ are parallel or anti-parallel (i.e., $\overline{R}_\times=0$), Eqs. (\ref{eq:nontrivial}) simplify to $\mu_1=\mu_2=\mu_3=0$ and $s_1=1$. This implies that $\bm{R}_i^{\prime\prime}=\bm{x}$ for $i=1,2$, or equivalently, that $\mathcal{D}$ outputs $\ket{+}=(\ket{0}+\ket{1})/\sqrt{2}$ independently of the input.

\paragraph*{Step 3.} For $\overline{R}_\times\neq 0$, the unitary $U$ rotates the input vectors $\bm{R}_i^{\prime\prime}$ to lie on the plane determined by the target vectors, and within that plane by a suitable angle. For $\overline{R}_\times = 0$, $U$ simply rotates the vector $\bm{x}$ in order to align it with $\overline{\bm{R}}_+$. In either case, $U$ can be expressed as the following map

\begin{equation}\label{eq:U}
\bm{R}_i^{\prime\prime}\stackrel{U}{\longmapsto}\bm{R}_i^{\prime\prime\prime}=k_{i1}\overline{\bm{R}}_1+k_{i2}\overline{\bm{R}}_2\,,
\end{equation}
with
\begin{align}
k_{ij}&=\tfrac{1}{\Gamma}\left[\alpha^2+\beta_i\beta_j\overline{R}_\times^2+(-1)^{i+j}\alpha\left(\beta_1-\beta_2\right)\overline{\bm{R}}_{\widetilde{i}}\cdot\overline{\bm{R}}_{\widetilde{j}}\right]\,,\\
\Gamma&=\sqrt{\alpha^2\overline{R}_+^2+\left[\|\beta_1\overline{\bm{R}}_1+\beta_2\overline{\bm{R}}_2\|^2+2\alpha\left(\beta_1-\beta_2\right)\right]\overline{R}_\times^2}\,.\label{eq:Gamma}
\end{align}
where we have defined the involutory map $\sim$ such that $\widetilde{1}=2$ and $\widetilde{2}=1$; and

\begin{align}
\alpha&=\bm{R}_i^{\prime\prime}\cdot \bm{x}=\sqrt{\frac{S+T}{2S}}\,,\label{eq:alpha_nt}\\
\beta_i&=\frac{\bm{R}_i^{\prime\prime}\cdot \bm{z}}{\overline{R}_\times}= \sqrt{\frac{2}{S(S+T)}} \bm{R}_i\cdot\bm{R}_-\,.\label{eq:beta_nt}
\end{align}

With this, the figure-of-merit of Eq. (\ref{eq:J}) can be shown to be
\begin{equation}
\mathcal{F}_{\rm HS}^{\rm A}=\frac{1}{2}+\frac{\Gamma_a}{2}\,,\label{eq:fid_nt}
\end{equation}
where $\Gamma_a$ is obtained by substituting Eqs. (\ref{eq:alpha_nt}) and (\ref{eq:beta_nt}) into Eq. (\ref{eq:Gamma}) and reads
\begin{equation}\label{eq:gamma_a}
\Gamma_a=\sqrt{\overline{R}_+^2+\frac{2R_-^2\overline{R}_\times^2}{S+T}}\,.
\end{equation}

\begin{figure*}
\begin{center}
\includegraphics[width=\textwidth]{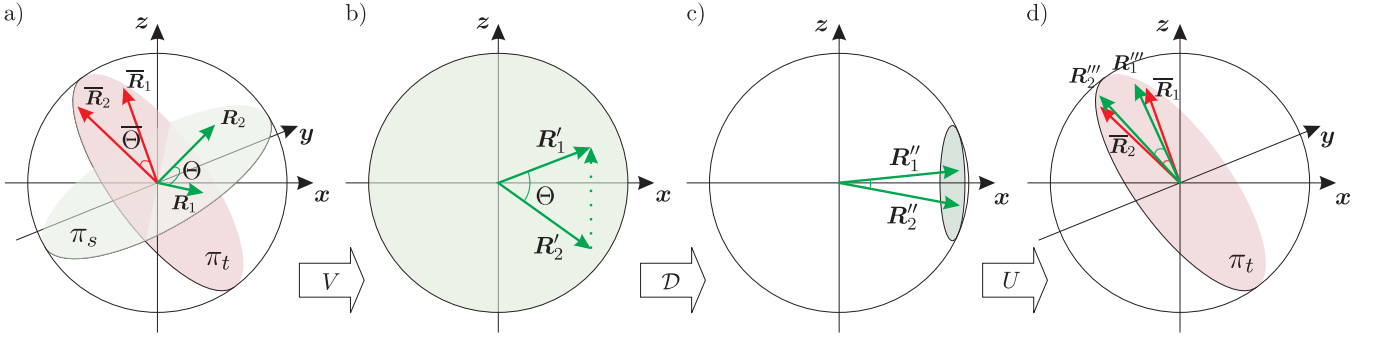}
\end{center}
\caption{(Color online) Bloch sphere schematics of procedure A.  a) The source (target) Bloch vectors $\bm{R}_1$ and $\bm{R}_2$ ($\overline{\bm{R}}_1$ and $\overline{\bm{R}}_2$) determine the plane $\pi_s$ ($\pi_t$). b) $V$ implements the rotation transforming $\pi_s$ to the $xz$-plane, in such a way that the vector $\bm{R}_1^\prime-\bm{R}_2^\prime$ is \emph{parallel} to $+\bm{z}$. c) The map $\mathcal{D}$ deforms the Bloch sphere into an ellipsoid of semi-axis $\mu_1$, $\mu_2$ and $\mu_3$ and translates it by $s_1$ along the $\bm{x}$ axis. The resulting ellipsoid touches the original Bloch sphere --- a feature related to the fact that $\mathcal{D}$ is an extremal CPTP map. d) $U$ rotates the resulting states to the plane $\pi_t$, and within that plane by some angle such that the resulting states $\bm{R}_1^{\prime\prime\prime}$ and $\bm{R}_2^{\prime\prime\prime}$ approximate $\overline{\bm{R}}_1$ and $\overline{\bm{R}}_2$, respectively.}\label{fig:solution}
\end{figure*}

Fig. \ref{fig:solution} illustrates this sequence of transformations for the case $\overline{R}_\times\neq 0$.

\subsubsection{Procedure B}\label{sec:procB}

As pointed out before, if $\Omega \leq 0$ then the affine transformation $\mathcal{D}$ is not implemented and the product $VU$ gives the unitary dynamics. In this case, $V$ can be chosen precisely as in step $1$ of procedure A.\par

  For $\overline{R}_\times\neq 0$, we preserve the form of the transformation $U$ from Eq. (\ref{eq:U}),
  but the values of $\alpha$ and $\beta_i$ are given by
\begin{align}
\alpha&=\bm{R}_i^{\prime}\cdot \bm{x}=\frac{R_\times}{R_-}\,,\label{eq:alpha_t}\\
\beta_i&=\frac{\bm{R}_i^{\prime}\cdot \bm{z}}{\overline{R}_\times}=\frac{\bm{R}_i\cdot\bm{R}_-}{\overline{R}_\times R_-}\,.\label{eq:beta_t}
\end{align}

For $\overline{R}_\times=0$, assume that the Bloch vectors $\overline{\bm{R}}_1$ and $\overline{\bm{R}}_2$ are anti-parallel (this is without loss of generality, since parallel targets always exhibit $\Omega > 0$ \footnote{To see this, note that if the targets are parallel it is immediate that $\Omega=|T|+T$ and our claims holds if $T>0$. A straightforward computation shows that $T=(\overline{R}_1+\overline{R}_2)^2 - (R_1^2\overline{R}_1^2 +R_2^2\overline{R}_2^2 + 2 R_1 R_2 \overline{R}_1\overline{R}_2 \cos{\Theta})$, which clearly achieves the minimum value $0$ iff $R_1=R_2=\cos{\Theta}=1$. However, this requires the sources to be identical, which is excluded from the statement of the problem \ref{problem}. Therefore, $T>0$.}). In particular, take $\overline{\bm{R}}_1$ parallel to $\bm{z}$ and $\overline{\bm{R}}_2$ parallel to $-\bm{z}$. The following transformation specifies $U$ in this case

\begin{multline}\label{eq:ULD}
\bm{R}_i^{\prime}\stackrel{U}{\longmapsto}\bm{R}_i^{\prime\prime}=\left[\frac{R_\times}{R_-}\cos{\vartheta}-\frac{(\bm{R}_i\cdot\bm{R}_-)}{R_-}\sin{\vartheta}\right]\bm{x}\\
+\left[\frac{R_\times}{R_-}\sin{\vartheta}+\frac{(\bm{R}_i\cdot\bm{R}_-)}{R_-}\cos{\vartheta}\right]\bm{z}\,,
\end{multline}
where
\begin{equation}\label{eq:sintheta}
\sin{\vartheta}=\frac{R_\times(\overline{R}_1-\overline{R}_2)}{R_-\sqrt{\overline{R}_+^2-T}}\,,
\end{equation}
and $\cos{\vartheta}=+\sqrt{1-\sin^2{\vartheta}}$. We note that $\vartheta$ is a valid angle since the rhs of Eq. (\ref{eq:sintheta}) is bounded between $-1$ and $1$ \footnote{This follows easily from the inequality $R_\times<R_-$ (Appendix \ref{app:ST}) and from $\overline{R}_+=|\overline{R}_1-\overline{R}_2|\leq \sqrt{\overline{R}_+^2-T}$ (the equality follows from the anti-parallelism of the target Bloch vectors and the inequality from the fact that $T\leq 0$ for $\overline{R}_\times = 0$ and $\Omega\leq 0$)}.

If $\overline{\bm{R}}_1$ and $\overline{\bm{R}}_2$ do not align along the $\bm{z}$ direction as specified above, we simply apply a further rotation that aligns the $\bm{z}$ axis with the direction $\overline{\bm{R}}_1/\overline{R}_1$.

 In both $\overline{R}_\times\neq 0$ and $\overline{R}_\times = 0$ cases, the average fidelity can be computed to be
\begin{equation}\label{eq:fid_t}
\mathcal{F}_{\rm HS}^{\rm B}=\frac{1}{2}+\frac{\Gamma_b}{2}\,,
\end{equation}
where $\Gamma_b$ is obtained by substituting Eqs. (\ref{eq:alpha_t}) and (\ref{eq:beta_t}) into Eq. (\ref{eq:Gamma}). After some manipulation we find
\begin{equation}
\Gamma_b\mathrel{\mathop:}=\sqrt{\overline{R}_+^2-T+2R_\times\overline{R}_\times}\,.
\end{equation}

\section{Optimality Proof}\label{sec:optimal}

In this section, we employ duality theory for SDPs to
prove the following theorem
\begin{theorem} \label{teo:optimality}
Our tracking strategy (as described in Sec. \ref{sec:strategy}), implements optimal tracking between any pair of source and target qubit states and is, therefore, a solution of the tracking problem introduced in Sec. \ref{sec:problem}.
\end{theorem}

In the subsequent proof of this theorem, some familiarity with SDP theory is assumed. Standard reviews on the topic are \cite{96Vandenberghe49,04boyd}. More closely related to our purposes is \cite{02Audenaert030302}, where the connection between optimization of quantum operations and SDPs was first noted. Also relevant is Ref. \cite{07Branczyk012329}, where a similar technique was used to approach a particular case of tracking.

\subsection{The tracking problem as a SDP}
We start by showing that the tracking problem can be formulated as a SDP. Formally, it can be written as
\begin{equation}\label{eq:opt1_app}
\max_{\mathcal{C}\in\{CPTP\}}{\sum_{i=1}^2{{\rm Tr}\left[\mathcal{C}(\rho_i)\pi_i\overline{\rho}_i\right]}}\,.
\end{equation}
It will be convenient to rewrite $\mathcal{C}(\rho_i)$ as \cite{01DAriano042308}
\begin{equation}\label{eq:ck}
\mathcal{C}(\rho_i)={\rm Tr}_A\left[(\rho_i^{\sf T}\otimes\openone)K_\mathcal{C}\right]\,,
\end{equation}
where $K_\mathcal{C}$ is the (unnormalized) Choi matrix \cite{75Choi285}
\begin{equation}\label{eq:kc}
K_\mathcal{C}=(\mathcal{I}\otimes\mathcal{C})(\ket{\Psi^+}\!\bra{\Psi^+})\,,
\end{equation}
and $\ket{\Psi^+}=\ket{00}+\ket{11}$. Eqs. (\ref{eq:ck}) and (\ref{eq:kc}) establish a one-to-one relation between the set of CPTP maps on qubits and the set of (unnormalized) 2-qubit density matrices satisfying ${\rm Tr_B} K_\mathcal{C}=\openone$ \cite{99Horodecki1888,99Fujiwara3290,01DAriano042308,75Choi285}. Here, ${\rm Tr}_{A(B)}$ denotes the partial trace operation over the first (second) qubit.\par

Using this isomorphism, a straightforward manipulation gives for the objective function in (\ref{eq:opt1_app}) the form $-{\rm Tr}\left[F_0 K_\mathcal{C}\right]$, where
\begin{equation}
F_0=-\sum_{i=1}^2{\rho_i^{\sf T}\otimes\pi_i\overline{\rho}_i}\,.
\end{equation}
whereas the constraint $\mathcal{C} \in \{CPTP\}$ becomes $K_\mathcal{C}\geq 0$ and ${\rm Tr}_B\left[K_\mathcal{C}\right]=\openone_A$. In conclusion, the tracking problem assumes the standard form of a SDP
\begin{equation}\label{eq:cvxopt}
  \begin{array}{rl}
  \text{maximize}&-{\rm Tr}\left[F_0 K_\mathcal{C}\right]\\
  \text{subject to}&K_\mathcal{C}\geq 0\\
  &{\rm Tr}\left[(\sigma_k\otimes\openone)K_\mathcal{C}\right] = 2\delta_{k 0}\,,\quad k=0,1,2,3.\\
  \end{array}
\end{equation}\par
 A special feature of (\ref{eq:cvxopt}) which will be explored next is that the replacement $\rho_i\mapsto A\rho_i A^\dagger$ and $\overline{\rho}_i \mapsto B^\dagger \overline{\rho}_i B$, with $A,B \in U(2)$, yields another SDP (with $\widetilde{F}_0$ replacing $F_0$) that achieves exactly the same optimal value. This can be easily seen by noting that: (i) ${\rm Tr}\left[F_0 K_\mathcal{C}\right]={\rm Tr}\left[\widetilde{F}_0 \widetilde{K}_\mathcal{C}\right]$, for $\widetilde{K}_\mathcal{C}=(A^{\sf T}\otimes B)^\dagger K_\mathcal{C} (A^{\sf T}\otimes B)$ and (ii) $\widetilde{K}_\mathcal{C}$ satisfies the constraints in (\ref{eq:cvxopt}) if and only if $K_\mathcal{C}$ does.\par

\subsection{The duality trick}
In terms of the Choi matrix, the strategy described in section \ref{sec:strategy} is written as $K_\mathcal{C}=(V^{\sf T}\otimes U) K_\mathcal{D} (V^{\sf T}\otimes U)^\dagger$, with
\begin{equation}
K_\mathcal{D}=\frac{1}{2}\openone\otimes\openone+\frac{s_1}{2}\openone\otimes X+\frac{\mu_1}{2}X\otimes X - \frac{\mu_2}{2}Y\otimes Y + \frac{\mu_3}{2}Z\otimes Z\,.
\end{equation}
Given the reasoning of the previous section, our strategy constitutes an optimal solution to the tracking problem if and only if the following SDP is solved with $\widetilde{K}=K_\mathcal{D}$,

\begin{equation}\label{eq:cvxopt_tilde}
  \begin{array}{rl}
  \text{maximize}&-{\rm Tr}\left[\widetilde{F}_0 \widetilde{K}\right]\\
  \text{subject to}&\widetilde{K}\geq 0\\
  &{\rm Tr}\left[(\sigma_k\otimes\openone)\widetilde{K}\right] = 2\delta_{k 0}\,,\quad k=0,1,2,3\,,\\
  \end{array}
\end{equation}
with
\begin{equation}
\widetilde{F}_0=-\sum_{i=1}^2{(V \rho_i V^\dagger)^{\sf T}\otimes U^\dagger\pi_i\overline{\rho}_i U}\,.
\end{equation}
The above SDP has the strong duality property, i.e., its optimal value is guaranteed to be identical to the optimal value of its dual problem \cite{04boyd}. This fact follows, for example, from the ``strict feasibility'' of the point $\widetilde{K}_f=\openone\otimes\openone/2$, which satisfies the constraints of (\ref{eq:cvxopt_tilde}) with the strict inequality $\widetilde{K}_f>0$.\par

From duality theory for SDPs, the problem above is solved with $\widetilde{K}=K_\mathcal{D}$ if and only if (i) $K_\mathcal{D}$ satisfies the constraints of  (\ref{eq:cvxopt_tilde}) and (ii) the linear matrix inequality
\begin{equation}\label{eq:lmi}
F=\widetilde{F}_0 + \mathfrak{x}_0 \openone\otimes \openone + \mathfrak{x}_1 X\otimes\openone + \mathfrak{x}_2 Y\otimes\openone + \mathfrak{x}_3 Z\otimes\openone \geq 0
\end{equation}
is satisfied by some quadruple $(\mathfrak{x}_0,\mathfrak{x}_1,\mathfrak{x}_2,\mathfrak{x}_3)$ such that
\begin{equation}\label{eq:weak_dual}
2 \mathfrak{x}_0=-{\rm Tr}\left[\widetilde{F}_0 K_\mathcal{D}\right]\,.
\end{equation}
 If that is the case, then the so-called ``complementary slackness'' condition \cite{02Audenaert030302}, $K_\mathcal{D}F=0$, holds for the appropriate values of coefficients $\mathfrak{x}_0$, $\mathfrak{x}_1$, $\mathfrak{x}_2$ and $\mathfrak{x}_3$.\par

To see that (i) is verified, recall that the values of $\mu_{1,2,3}$ and $s_1$ were chosen to make of $\mathcal{D}$ an (extreme) CPTP map. As mentioned before, the Choi matrix of any such map (on qubits) is characterized by the constraints of problem (\ref{eq:cvxopt_tilde}).\par

For (ii), first note that
$-{\rm Tr}\left[\widetilde{F}_0 K_\mathcal{D}\right]$ is merely $\mathcal{F}_{\rm HS}^{\rm A}$ or $\mathcal{F}_{\rm HS}^{\rm B}$ given in Eqs. (\ref{eq:fid_nt}) and (\ref{eq:fid_t}), depending on whether $\Omega>0$ or $\Omega \leq 0$. In our particular problem, the complementary slackness condition results in sufficient independent linear equations that $\mathfrak{x}_0$, $\mathfrak{x}_1$, $\mathfrak{x}_2$ and $\mathfrak{x}_3$ are defined precisely. We find $\mathfrak{x}_2=0$ and

\begin{itemize}
\item If $\Omega>0$,
\begin{subequations}\label{eq:coefomegap}
\begin{align}
\mathfrak{x}_0&=\frac{1}{4}\left(1+\Gamma_a\right)\,,\\
\mathfrak{x}_1&=\frac{R_\times}{4R_-}\left(1+\Gamma_a\right)\,,\\
\mathfrak{x}_3&=\frac{1}{4R_-}\left[\left(\pi_1\bm{R}_1+\pi_2\bm{R}_2\right)\cdot\bm{R}_-+\frac{\Xi}{\Gamma_a}\right]\,. \label{eq:xi3p}
\end{align}
\end{subequations}
\item If $\Omega\leq 0$,
\begin{subequations}\label{eq:coefomegam}
\begin{align}
\mathfrak{x}_0&=\frac{1}{4}\left(1+\Gamma_b\right)\,,\\
\mathfrak{x}_1&=\frac{1}{4R_-}\left( R_\times+\frac{\xi}{\Gamma_b}\right)\,,\label{eq:xi1m}\\
\mathfrak{x}_3&=\frac{1}{4R_-}\left[\left(\pi_1\bm{R}_1+\pi_2\bm{R}_2\right)\cdot\bm{R}_-+\frac{\Xi}{\Gamma_b}\right]\,.\label{eq:xi3m}
\end{align}
\end{subequations}
\end{itemize}
where, for brevity, we have defined
\begin{align}
\Xi&\mathrel{\mathop:}= \sum_{i=1}^2{\left(\bm{R}_i\cdot\bm{R}_-\right)\left(\overline{\bm{R}}_i\cdot\overline{\bm{R}}_+\right)}\,,\\
\xi&\mathrel{\mathop:}= R_\times\overline{R}_+^2+\overline{R}_\times R_-^2\label{eq:xi}\,,
\end{align}
which can be shown to satisfy the relation
\begin{equation}\label{eq:xixi}
\Xi^2+\xi^2=R_-^2\overline{R}_+^2\Gamma_b^2\,.
\end{equation}
We prove in Appendix \ref{app:welldef} that, although $\Gamma_{a}$ and $\Gamma_b$ appear in the denominator of some of the coefficients in Eqs. (\ref{eq:coefomegap}) and (\ref{eq:coefomegap}), no singularities occur if the indicated range of $\Omega$ is observed.

With the set of coefficients (\ref{eq:coefomegap}) and (\ref{eq:coefomegam}), Eq. (\ref{eq:weak_dual}) is clearly satisfied. As a result, the optimality of our tracking strategy is solely dependent on proving the linear matrix inequality $F\geq 0$ for the above set of coefficients. In Appendix \ref{app:charpoly} we study the characteristic polynomial of $F$ and conclude that all of its roots are non-negative, thus proving Theorem \ref{teo:optimality}.

\section{Examples}\label{sec:examples}
In this section we evaluate our tracking strategy at work in some physically relevant problems such as quantum state discrimination, quantum state purification, stabilization of quantum states in the presence of noise and state-dependent quantum cloning. Moreover, we also discuss the application of our strategy in circumstances where tracking is known to be perfectly achievable. The analysis presented in this section is meant to give an explicit account on the wide range of physical applications of the tracking problem and its optimal solution.

 \subsection{Quantum State Discrimination}\label{sec:ex1}
A standard result in quantum state discrimination is the Helstrom measurement \cite{76Helstrom}, which consists of a projective quantum measurement that maximizes the probability ($P_{\rm Helst}$) of correctly identifying the state of a quantum system that could have been prepared in two different states. Describing the possible preparations by $\rho_1$ with probability $p_1$ and $\rho_2$ with probability $p_2$, the Helstrom measurement gives
\begin{equation}\label{eq:helstrom_prob}
P_{\rm Helst}=\tfrac{1}{2}+\tfrac{1}{2}\|p_1\rho_1-p_2\rho_2\|_{\rm tr}\,,
\end{equation}
where $\|\cdot\|_{\rm tr}$ denotes the trace norm.\par
In this section, we propose a quantum state discrimination protocol for a pair of qubit states based on the tracking strategy introduced in the last section. We will show that it is equivalent to Helstrom's strategy, as it will give the same correct identification probability of Eq. (\ref{eq:helstrom_prob}).

Our quantum state discrimination protocol consists of two simple steps: first we apply the tracking operation $\mathcal{C}$
introduced in section \ref{sec:strategy} to approximate the states to be discriminated to some pair of orthogonal states. Without loss of generality, we take $\overline{\rho}_1=\ket{0}\!\bra{0}$ and $\overline{\rho}_2=\ket{1}\!\bra{1}$. The priority of each transformation is taken to be identical to the prior probabilities with which $\rho_1$ and $\rho_2$ are prepared, i.e., $\pi_i=p_i$. As the second and final step, we perform the quantum measurement $\{\ket{0}\!\bra{0},\ket{1}\!\bra{1}\}$, under the understanding that an outcome `0' suggests the preparation to be $\rho_1$ and an outcome `1' suggests $\rho_2$.\par
The probability of a correct identification under this tracking scheme is given by Born's rule, averaged with the prior probabilities,
\begin{align}
P_{\rm track}&=p_1{\rm Tr}\left[\mathcal{C}(\rho_1)\ket{0}\!\bra{0}\right]+p_2{\rm Tr}\left[\mathcal{C}(\rho_2)\ket{1}\!\bra{1}\right]\nonumber\\
&=\pi_1{\rm Tr}\left[\mathcal{C}(\rho_1)\overline{\rho}_1\right]+\pi_2{\rm Tr}\left[\mathcal{C}(\rho_2)\overline{\rho}_2\right]\,.\label{eq:acc_prob}
\end{align}
By comparing Eqs. (\ref{eq:acc_prob}) and (\ref{eq:J}), one promptly recognizes that $P_{\rm track}$ is precisely the performance of the operation $\mathcal{C}$ for tracking from $\rho_i$ to $\overline{\rho}_i$ with priority $\pi_i$, as measured by $\mathcal{F}_{\rm HS}$. Hence, in the case of $\overline{\rho}_1=\ket{0}\!\bra{0}$, $\overline{\rho}_2=\ket{1}\!\bra{1}$ and $\pi_i=p_i$, Eqs. (\ref{eq:fid_nt}) and (\ref{eq:fid_t}) give the probability of success of our discrimination scheme for $\Omega>0$ and $\Omega \leq 0$, respectively. Next, we make these formulas more explicit.\par

Using the condition $\overline{R}_\times=0$ in Eqs. (\ref{eq:S}) and (\ref{eq:IF}), we obtain $\Omega=|T|+T$. Essentially, this means that $T$
assumes the role of the indicator function: if $T>0$, then $\Omega>0$ and we employ procedure A; if $T\leq 0$, then $\Omega=0$ and we employ procedure B.
Substituting $\overline{\bm{R}}_i=-(-1)^i p_i\bm{z}$ into Eq. (\ref{eq:T}), some simple algebra gives
\begin{equation}
T=(p_1-p_2)^2-\|p_1\bm{R}_1-p_2\bm{R}_2\|^2\,,
\end{equation}
and we can write
\begin{equation}
P_{\rm track}=\left\{\begin{array}{rcl}
\tfrac{1}{2}+\tfrac{1}{2}|p_1-p_2|&\mbox{if}&T>0\\
\tfrac{1}{2}+\tfrac{1}{2}\|p_1\bm{R}_1-p_2\bm{R}_2\|&\mbox{if}&T\leq 0
\end{array}
\right.\,,
\end{equation}
where the first line follows from Eq. (\ref{eq:fid_nt}) and the second from Eq. (\ref{eq:fid_t}).\par

 It is a tedious exercise (essentially the computation of the eigenvalues of $p_1\rho_1-p_2\rho_2$)
 to re-express Eq. (\ref{eq:helstrom_prob}) in terms of the Bloch vectors $\bm{R}_i$. The result is exactly
\begin{equation}
P_{\rm Helst}=P_{\rm track}\,,
\end{equation}
hence establishing the claimed equivalence between our strategy and Helstrom's.

Note that if $T>0$, $P_{\rm track}$ is independent of the states we are trying to distinguish, but merely dependent on the probabilities with which they occur. This can be understood by looking at the details of the affine operation taking place in procedure A. As noted before, for $\overline{R}_\times=0$ (as is the case for orthogonal targets), the affine map is such that $\mu_1=\mu_2=\mu_3=0$ and $s_1=1$; that is, the source states are completely depolarized and a new state $\ket{+}$ is prepared instead. Next, this state is rotated by the unitary $U$ and the measurement is finally performed.

\begin{figure}
\includegraphics[width=8cm]{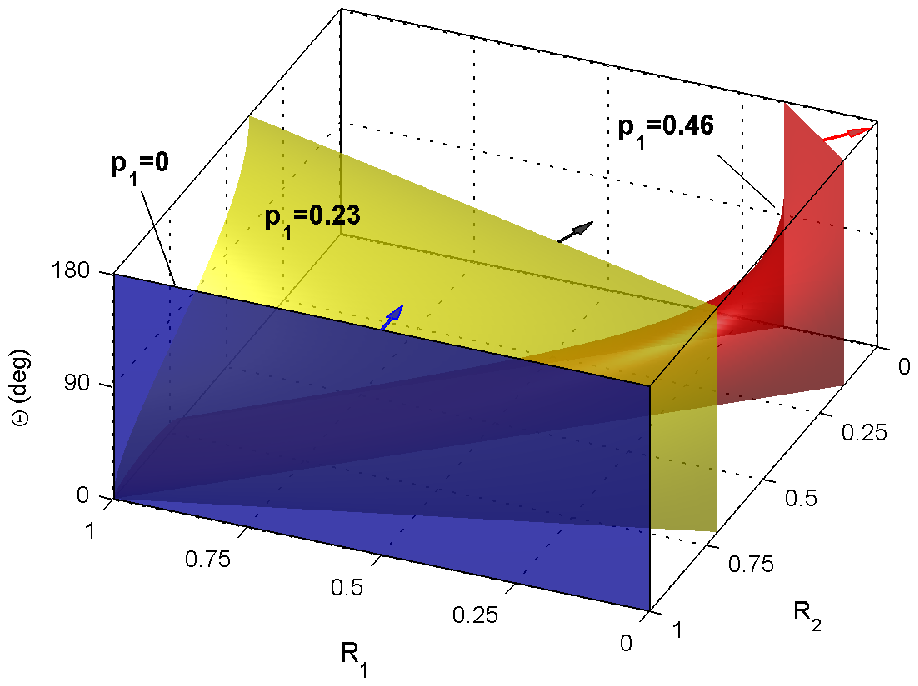}
\\
\includegraphics[width=8cm]{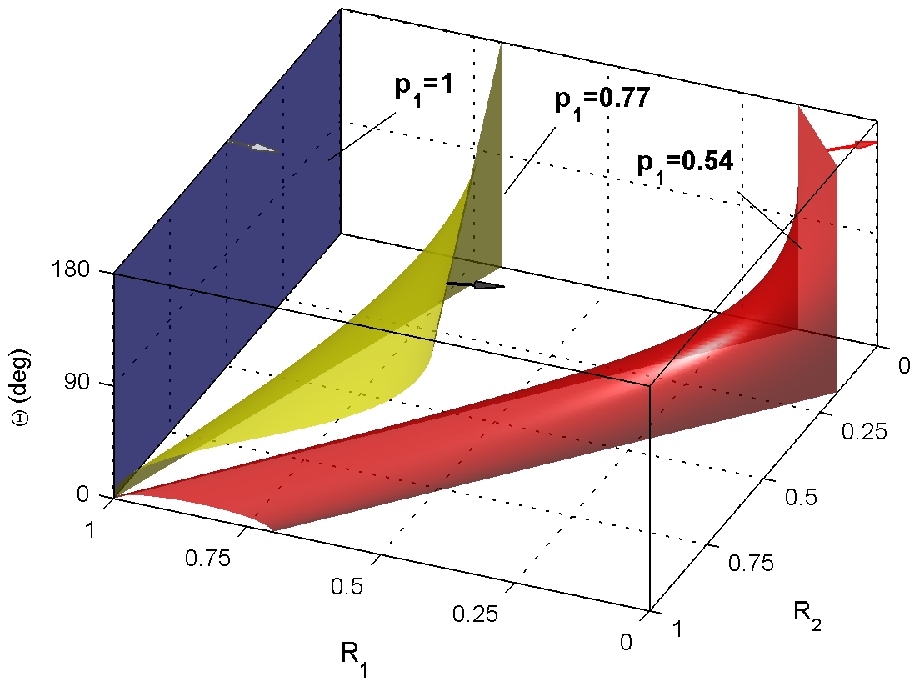}
\caption{(Color online) Each color of sheet represents a fixed deviation from the uniform probability distribution. The sheets divide the parameter space in two regions. The arrows designate the regions where procedure A (non-unitary) is recommended. On the other side, procedure B (unitary) is recommended. The larger the deviation from $p_1=1/2$, the larger the region where a non-unitary preparation for the measurement $\{\ket{0}\!\bra{0},\ket{1}\!\bra{1}\}$ is advisable.}\label{fig:beat_Helst}
\end{figure}

It is easy to see that for $p_1=p_2=1/2$ the condition $T > 0$ (procedure A) never holds. However, as we deviate from the uniform distribution, the volume of the parameter space where procedure A is recommended grows to fully cover the space when $p_1=0$ or $p_1=1$. This is shown in Fig. \ref{fig:beat_Helst} .

\subsection{Quantum state Purification}\label{sec:ex1p5}

In this section, we consider a kind of state purification task where we aim to transform a pair of mixed source states $\rho_1$ and $\rho_2$ with the same degree of mixedness ($R_1=R_2=R<1$) that are separated in the Bloch sphere by an angle $2\theta \in (0,\pi]$ into a pair of pure target states $\overline{\rho}_1$ and $\overline{\rho}_2$ separated by the same  Bloch sphere angle $2\theta$. In other words, our purification task consists of elongating the Bloch vectors while preserving the angle between them.

For later use, it will be convenient to derive formulas for the indicator function and figure-of-merit of a slightly more general problem, where the angle between the target Bloch vectors is $2\overline{\theta} \in [0,\pi]$. The purification task can be recovered by restricting $\theta=\overline{\theta}$. In addition, we will allow the priorities of the transformations $\rho_i\to\overline{\rho}_i$ to be arbitrary positive scalars $\pi_i$ such that $\pi_1+\pi_2=1$. Later, we make $\pi_1=\pi_2$, in order to simplify the formulas.\par

In this generalized purification framework ($\theta\neq\overline{\theta}$), the indicator function is obtained from Eq. (\ref{eq:IF}), by incorporating the conditions $\overline{R}_i=\pi_i$ (purity of the targets) and  $R_i=R$ (common mixedness of the sources) in the expressions for $R_\times\overline{R}_\times$, $T$ and $S$, from Eqs. (\ref{eq:Rx}), (\ref{eq:T}) and (\ref{eq:S}), respectively. These have a particularly appealing form:
\begin{subequations}\label{eq:RSTpurif}
\begin{align}
R_\times\overline{R}_\times&=\sqrt{\mathcal{R}_s\mathcal{R}_c\left[\Pi_+\Pi_--\delta\right]}\,,\\
T&=(1-\mathcal{R}_c)\Pi_+-\mathcal{R}_s\Pi_-\,,\\
S&=\sqrt{\left[(1-\mathcal{R}_c)\Pi_++\mathcal{R}_s\Pi_-\right]^2-4\delta\mathcal{R}_s(1-\mathcal{R}_c)}\,,
\end{align}
\end{subequations}
where we have defined $\mathcal{R}_c \mathrel{\mathop:}= R^2\cos^2\theta$, $\mathcal{R}_s\mathrel{\mathop:}=R^2\sin^2\theta$, $\delta=(\pi_1-\pi_2)^2$ and
\begin{equation}
\Pi_\pm\mathrel{\mathop:}=\pi_1^2+\pi_2^2\pm2\pi_1\pi_2\cos{2\overline{\theta}}\,.
\end{equation}
From the above equations, the indicator function and the figure-of-merit can be immediately obtained. At this point, though, we specialize to the case $\delta=0$ (i.e., $\pi_1=\pi_2=1/2)$ and give explicit formulas in this particular case. From Eq. (\ref{eq:IF}),
\begin{align}
\Omega&=2\left[(1-\mathcal{R}_c)\Pi_+-\sqrt{\mathcal{R}_s\mathcal{R}_c\Pi_+\Pi_-}\right]\\
&=2\left[\cos^2\overline{\theta}-R^2\cos{\theta}\cos{\overline{\theta}\cos{\left(\theta-\overline{\theta}\right)}}\right]\,,\label{eq:Omegapurif}
\end{align}
where, in the second line, we used that $\Pi_+=\cos^2\overline{\theta}$ and $\Pi_-=\sin^2\overline{\theta}$ when $\pi_1=\pi_2=1/2$. From Eqs. (\ref{eq:fid_nt}) and (\ref{eq:fid_t}),
\begin{equation}\label{eq:fidpurif}
\mathcal{F}_{\rm HS}=\left\{
\begin{array}{rcl}
\frac{1}{2}+\frac{1}{2}\sqrt{\cos^2\overline{\theta}+\frac{R^2\sin^2\theta\sin^2\overline{\theta}}{1-R^2\cos^2\theta}}&\mbox{if}&\Omega >0\\
\frac{1}{2}+\frac{1}{2}R\cos{\left(\theta-\overline{\theta}\right)}&\mbox{if}&\Omega \leq0
\end{array}\right.\,.
\end{equation}

\begin{figure}
\includegraphics[width=8.5cm]{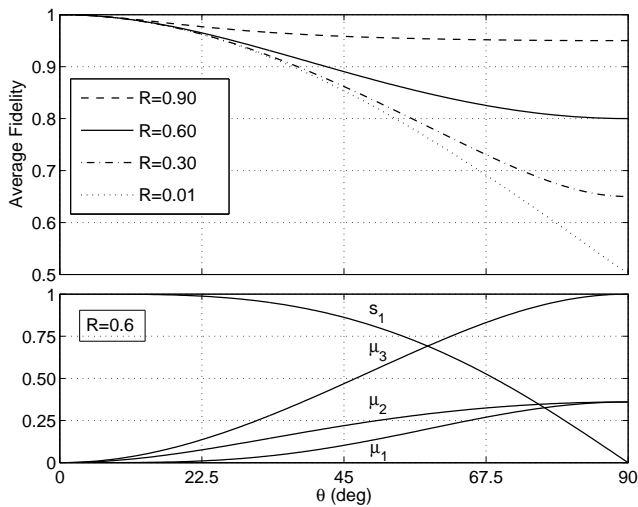}
\caption{Purifying a pair of mixed states with Bloch vectors of length $R<1$, separated by an angle $2\theta\in(0,\pi]$ with priorities $\pi_1=\pi_2=1/2$. The plot shows the optimal average fidelity for different values of $R$ and the control parameters $s_1$ and $\mu_{\{1,2,3\}}$ for $R=0.6$. The purification procedure attempts to increase, as much as possible, the length $R$ while preserving the angle $\theta$.}\label{fig:purif}
\end{figure}

It is now straightforward to see that, if $\theta=\overline{\theta}$, then $\Omega \geq 0$ with saturation if and only if $\theta = \pi/2$. That is, unless we are trying to purify from antipodal mixed states to orthogonal states, the best strategy is always a non-unitary transformation (procedure A). The optimal average fidelity of the purification scheme can be obtained by using $\theta=\overline{\theta}$ in Eq. (\ref{eq:fidpurif}). The resulting optimal purification performance is shown in Fig. \ref{fig:purif} and corresponds to the best achievable average fidelity allowed by quantum mechanics to the purification problem at hand.\par

From Fig. \ref{fig:purif}, we see that for small $\theta$, $\mathcal{F}_{\rm HS}$ is typically high, regardless of the length $R$. This can be understood in analogy to the fact that collapsing a set of mixed states into a single pure state is always perfectly achievable. In fact, such a collapse is nearly what is needed in this domain, since a pair of pure target states separated by a small angle can be well approximated by a single pure state. Fig. \ref{fig:purif} confirms this reasoning by showing that, in the small $\theta$ domain, the source Bloch vectors are strongly compressed due to the small values of $\mu_{\{1,2,3\}}$ and then strongly elongated due to the large value of $s_1$.\par

For increasing values of $\theta$ the fidelity decreases. Such a decay is accentuated if the degree of mixedness of the source states is high (small values of $R$), reflecting the intuitive idea that it is harder to purify very mixed states. In these intermediate regions, a non-trivial combination of compressions $\mu_{\{1,2,3\}}$ and translation $s_1$ of the Bloch vectors forms the optimal purifying scheme. Noticeably, the optimal procedure has less effect on the qubit (decreasing $s_1$ and increasing $\mu_{\{1,2,3\}}$) as $\theta$ increases.\par

At $\theta=\pi/2$, we have $\Omega=0$ and an optimal unitary transformation is actually to do nothing (the unitary transformation $V$ is undone by another unitary $V^\dagger$, see Sec. \ref{sec:procB}). Note that although the plot of $\mu_1$ and $\mu_2$ in Fig. \ref{fig:purif} approaches a constant value in between $0$ and $1$, the vanishing indicator function introduces a discontinuity in the purifying operation, since now we should use procedure B, hence $\mu_1=\mu_2=1$ at $\theta=\pi/2$. Nevertheless, the values of $\mu_1$ and $\mu_2$ are utterly irrelevant in this case. At this stage both Bloch vectors are aligned with the $\bm{z}$ direction, thus any compression along $\bm{x}$ and $\bm{y}$ cannot affect the states of interest.

\subsection{Stabilizing pure states}\label{sec:ex2}

A possible use for tracking is to try to cope with the presence of noise in quantum computation and communication involving qubits. In general, noise processes (we restrict ourselves to CP processes) cannot be inverted by another CP map, not even when the noise is perfectly known \footnote{This follows from the semi-group structure of CP maps. An obvious exception arises by restricting to the group of unitary noises. In fact, a theorem by Wigner states that this is the only exception (see \cite{05Buscemi082109} for a proof, see also \cite{07Nayak103})}. However, instead of stabilizing the full Bloch sphere against noise, one may be interested at stabilizing only a limited number of states. Although not perfect, it is not uncommon that good stabilization can be achieved within this framework.\par

In this section, we consider a quantum error correction task of this type, which was studied in detail in Ref. \cite{07Branczyk012329}. We will show that the optimal correction scheme is merely a particular case of the quantum state purification procedure (with $\theta\neq\overline{\theta}$) introduced in the previous section.\par

Assume that Alice prepares (with equal probabilities) a qubit in one of the non-orthogonal pure states
\begin{subequations}\label{eq:ag_states}
\begin{align}
\ket{\psi_1}&=\cos{\frac{\overline{\theta}}{2}}\ket{+}+\sin{\frac{\overline{\theta}}{2}}\ket{-}\,,\\
\ket{\psi_2}&=\cos{\frac{\overline{\theta}}{2}}\ket{+}-\sin{\frac{\overline{\theta}}{2}}\ket{-}\,,
\end{align}
\end{subequations}
where $\ket{\pm}=(\ket{0}\pm\ket{1})/\sqrt{2}$ and $\overline{\theta}$ is the half-angle between $\ket{\psi_1}$ and $\ket{\psi_2}$ in the Bloch sphere representation, hence $\overline{\theta}\in(0,\pi/2)$. She then sends her qubit to Bob through a dephasing channel
\begin{equation}\label{eq:dephasing}
\mathcal{E}(\varsigma)=p Z \varsigma Z+(1-p)\varsigma\,,
\end{equation}
where $p$ is a constant in the range $(0,1/2]$ that has been previously determined. Bob, who does not know which of the two states was prepared, has to apply a quantum operation so as to ensure that, when Alice performs a check-measurement $\{\ket{\psi_{k}}\!\bra{\psi_{k}},\openone-\ket{\psi_{k}}\!\bra{\psi_{k}}\}$ (with $k$ labeling the identity of her actual preparation) on Bob's output, the probability of detecting her original preparation is as high as possible. This probability equals the average fidelity between the possible inputs and the outputs of Bob's operation.\par

Our tracking strategy can be of assistance to Bob if he regards the two possible noisy states as the source states $\rho_i=\mathcal{E}(\ket{\psi_i}\!\bra{\psi_i})$ and tracks (with equal priorities $\pi_i=1/2$) to the target states $\overline{\rho}_i=\ket{\psi_i}\!\bra{\psi_i}$. In this case, the target states are pure and the source states have the same degree of mixedness [this follows easily from the application of the dephasing map to the states of Eq. (\ref{eq:ag_states})], which is precisely the scenario we considered in the last section for quantum state purification.

The indicator function $\Omega$ can then be obtained from Eq. (\ref{eq:Omegapurif}) by using the following identities for the angle $\theta$ (recall that $\theta$ is the half-angle, in the Bloch sphere, between the states output by the dephasing noise),
\begin{equation}\label{eq:dephident}
\sin{\theta}=\frac{\sin{\overline{\theta}}}{R}\quad\mbox{and}\quad\cos\theta=\frac{(1-2p)\cos{\overline{\theta}}}{R}\,,
\end{equation}
where $R$ is the length of the noisy Bloch vectors. Explicitly,
\begin{equation}
\Omega=2\cos^2\overline{\theta}\left[1-R^2+2p\sin^2\overline{\theta}\right]\,.
\end{equation}
It is easy to see that, given the ranges $\overline{\theta} \in (0,\pi/2)$ and $p\in(0,1/2]$, we have $\Omega > 0$, which implies that Bob should always apply the non-unitary procedure A. The optimal performance is then obtained by substituting the identities (\ref{eq:dephident}) in the first line of Eq. (\ref{eq:fidpurif}), which gives

\begin{equation}\label{eq:fid_aggie}
\mathcal{F}_{\rm HS}=\frac{1}{2}+\frac{1}{2}\sqrt{\cos^2{\theta}+\frac{\sin^4{\theta}}{1-\left(1-2p\right)^2\cos^2{\theta}}}\,.
\end{equation}
As expected, this is precisely the optimal fidelity found for this problem  in \cite{07Branczyk012329}.\par

It should be clear that our tracking strategy can be similarly applied to the stabilization of quantum states different from those of Eq. (\ref{eq:ag_states}), prepared with non-uniform prior probabilities and undergoing noise dynamics different from dephasing, in any case still providing optimal stabilization. It thus represents a significant extension of the results in \cite{07Branczyk012329}.

\subsection{Perfectly tracking quantum states}\label{sec:ex3}
In this section we evaluate the performance of our strategy in circumstances where tracking is known to be perfectly achievable. It will be convenient to split our analysis in two, namely, the case of two pure target states and the remaining cases (in which at least one of the target states is mixed).

\subsubsection{Pure target states}\label{sec:ex31}
In appendix \ref{app:ptc} we prove a corollary of Alberti and Uhlmann's theorem stating that a CPTP  map $\mathcal{A}$ perfectly transforming a pair of quantum states $\rho_i$ ($i=1,2$) into a pair of pure states $\overline{\rho}_i$ exists if and only if $\rho_i$ are also pure and $\theta\geq\overline{\theta}$. Since our tracking strategy is  optimal (cf. Theorem \ref{teo:optimality}), we can infer from Alberti and Uhlmann's theorem that it implements tracking with unit fidelity whenever $R=1$ and $\theta\geq\overline{\theta}$. This is explicitly verified in the sequence, where the indicator function and the figure-of-merit for pure state transformations are computed.

We start using Eq. (\ref{eq:RSTpurif}) with $R=1$ (pure source condition) to construct the indicator function $\Omega$ from Eq. (\ref{eq:IF}). After some straightforward manipulation, we obtain
\begin{equation}
\Omega=8\pi_1\pi_2\sin{\theta}\cos{\overline{\theta}}\sin{\left(\theta-\overline{\theta}\right)}\,.
\end{equation}
For our purposes, the only meaningful feature of $\Omega$ is whether it is strictly positive or not, in which case the above expression is equivalent to
\begin{equation}\label{eq:omegaequiv}
\widetilde{\Omega}=2\left(\theta-\overline{\theta}\right)\,,
\end{equation}
since $\theta,\overline{\theta}\in(0,\pi/2]$ and $\pi_1,\pi_2 \in (0,1)$. Recall that $\widetilde{\Omega}$ is the indicator function obtained in Sec. \ref{sec:indicator}, Eq. (\ref{eq:IF_simple}), via an heuristic argument.\par
The figure-of-merit, in turn, can be obtained from Eqs. (\ref{eq:fid_nt}) and (\ref{eq:fid_t}) to be $\mathcal{F}_{\rm HS}^{\rm A} = 1$ (if $\widetilde{\Omega}>0$) and
\begin{equation}\label{eq:fidppunit}
\mathcal{F}_{\rm HS}^{\rm B} =\frac{1}{2}+\frac{1}{2}\sqrt{\pi_1^2+\pi_2^2+2\pi_1\pi_2\cos{\left(2\theta-2\overline{\theta}\right)}}
\end{equation}
(if $\widetilde{\Omega} \leq 0$). Note, however, that $\mathcal{F}_{\rm HS}^{\rm B}=1$ if $\theta=\overline{\theta}$ (i.e., $\widetilde{\Omega}=0$), in such a way that we can write

\begin{equation}\label{eq:fid_B}
\mathcal{F}_{\rm HS}=\left\{
\begin{array}{ccl}
1&\mbox{if}&\theta\geq\overline{\theta}\\
\mathcal{F}_{\rm HS}^{\rm B}&\mbox{if}&\theta<\overline{\theta}
\end{array}
\right.\,.
\end{equation}

The first line of Eq. (\ref{eq:fid_B}) is exactly the content of Alberti and Uhlmann's theorem applied to pure state transformations, whereas the second line establishes the optimal achievable average fidelity when perfect pure state transformation is impossible.

In conclusion, besides representing a construction of Alberti and Uhlmann's map $\mathcal{A}$ for perfect pure state transformations, our tracking strategy also gives the unitary map (procedure B) that optimally approximates impossible pure state transformations.

\subsubsection{Mixed target states}
The requirement of perfect tracking does not restrict the target states to be pure. In fact, the more general form of Alberti and Uhlmann's theorem states that for any given target states $\overline{\rho}_i$, there exists a CPTP map $\mathcal{A}$ that implements perfect tracking from all source states $\rho_i$ satisfying
\begin{equation}\label{eq:AU}
\|\overline{\rho}_1-t\overline{\rho}_2\|_{\rm tr}\leq\|\rho_1-t\rho_2\|_{\rm tr}\quad\forall t\in\mathbb{R}^+\,,
\end{equation}
In contrast to the previous section though, our tracking strategy is generally not a construction of the map $\mathcal{A}$ in this case. As mentioned before, this is a consequence of the fact that our figure-of-merit is not as well motivated in the case of mixed target states. For example,
in situations where perfect tracking is possible, the resulting average Hilbert-Schmidt inner product does not achieve its maximal value. This is further explored next.\par

\begin{figure}
\includegraphics[width=8.5cm]{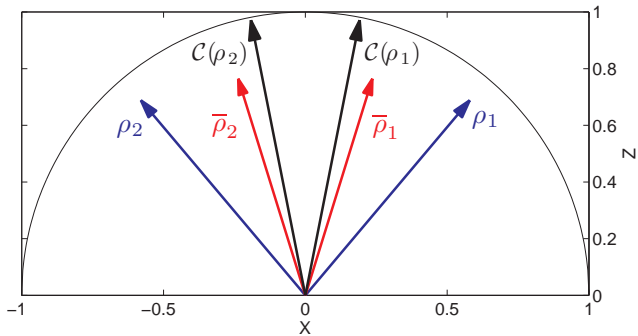}
\caption{(Color online) Although perfect tracking $\rho_i\to\overline{\rho}_i$ is physically allowed, the resulting average Hilbert-Schmidt inner product [the rhs of Eq. (\ref{eq:perfcond})] is only $0.82$ for this transformation. Our tracking strategy $\mathcal{C}$ attempts to maximize this number, finding a different transformation which gives an average Hilbert-Schmidt inner product [the lhs of Eq. (\ref{eq:perfcond})] equal to approximately $0.89$. As a result, $\mathcal{C}$ does not implement perfect tracking.}\label{fig:mixtrack}
\end{figure}

Any CPTP map $\mathcal{C}$ implementing perfect tracking must satisfy
\begin{equation}\label{eq:perfcond}
\pi_1{\rm Tr}\left[\mathcal{C}(\rho_1)\overline{\rho}_1\right]+\pi_2{\rm Tr}\left[\mathcal{C}(\rho_2)\overline{\rho}_2\right]=\pi_1{\rm Tr}\overline{\rho}_1^2+\pi_2{\rm Tr}\overline{\rho}_2^2\,.
\end{equation}
Our strategy, though, does not arise from an attempt to enforce Eq. (\ref{eq:perfcond}), but instead to maximize its lhs (cf. section \ref{problem}). Although these actions are equivalent in the case of pure target states [the rhs of Eq. (\ref{eq:perfcond}) equals $1$, which is precisely the maximum value of its lhs for states satisfying the criterion of Eq. (\ref{eq:AU})], for mixed target states this equivalence is lost. In this case, the lhs can typically be made greater than the rhs by employing an operation $\mathcal{C}$ that elongates the source Bloch vectors to nearly pure states, as illustrated in Fig. \ref{fig:mixtrack}. As a consequence, the maximization of our figure-of-merit leads to a departure from the perfect tracking operation.

Yet, recall that the average Hilbert-Schmidt inner product lower bounds the average fidelity and as such, its maximization has some beneficial impact in implementing tracking, in the sense that it ensures that the resulting average fidelity is no less than the maximal average Hilbert-Schmidt inner product.

\subsection{State-dependent Cloner}

One of the most celebrated results in quantum information science is the ``no-cloning theorem'' \cite{82Dieks271,82Wootters802}, which establishes the impossibility of copying an unknown pure quantum state. Since its inception in the literature, a lot of work has been done on the topic, both extending its range of applicability as well as attempting to weaken its impact in practical applications (see \cite{05Scarani1225} for a review). Remarkable results in each of these directions are the ``no-broadcasting theorem'' for noncommuting mixed quantum states \cite{96Barnum2818} and the Bu\v{z}ek-Hillery optimal quantum cloning machine \cite{96Buzek1844}. \par

In this section we consider a state-dependent cloning task introduced in Ref. \cite{98Bruss2368}. We will show that our tracking strategy provides a straightforward derivation of the optimal cloning fidelity obtained in that paper. Following \cite{98Bruss2368}, let
\begin{subequations}
 \begin{align}
\ket{a}&=\cos{\phi}\ket{0}+\sin{\phi}\ket{1}\,,\\
\ket{b}&=\sin{\phi}\ket{0}+\cos{\phi}\ket{1}\,,
 \end{align}
 \end{subequations}
 for $\phi\in [0,\pi/4)$, be the only two possible preparations of a single-qubit, each of which occurring with probability $1/2$. The cloning task is to output the two-qubit state $\ket{aa}\equiv\ket{a}\otimes\ket{a}$ if the initial preparation is $\ket{a}$ or $\ket{bb}\equiv\ket{b}\otimes\ket{b}$ if the initial preparation is $\ket{b}$. In \cite{98Bruss2368}, a unitary transformation $U$ was obtained such that the figure-of-merit (the so-called ``global fidelity'')
\begin{equation}\label{eq:Fg}
F_g=\frac{1}{2}\left(|\langle a a |U| a 0\rangle|^2+|\langle bb |U| b 0\rangle|^2\right)
\end{equation}
is maximal.\par

The key point that allows the application of our tracking strategy here is that, although the unitary evolution $U$ acts on the Hilbert space of a two-qubit system, it was shown in \cite[Appendix B]{98Bruss2368} that the maximizing $U$ is such that $U\ket{a0}$ and $U\ket{b0}$ lie in the two-dimensional subspace spanned by $\{\ket{aa},\ket{bb}\}$. Therefore, we can regard this cloning as a transformation from the two-dimensional subspace spanned by $\{\ket{a0},\ket{b0}\}$ to the two-dimensional subspace spanned by $\{\ket{aa},\ket{bb}\}$. By this same argument, we could have even relaxed the condition that the system to be cloned is a qubit.

Let $\ket{s_1}$ and $\ket{s_2}$ ($\ket{t_1}$ and $\ket{t_2}$) be the fictitious qubit source (target) states, and let $2\theta$ ($2\overline{\theta}$) be the Bloch sphere angle between them. Then, we must have
\begin{subequations}
\begin{align}
\langle s_1 | s_2 \rangle &= \langle a0 | b0 \rangle = \sin{(2\phi)} = \cos\theta\,,\\
\langle t_1 | t_2 \rangle &= \langle aa | bb \rangle = \sin^2{(2\phi)} = \cos\overline{\theta}\,.
\end{align}
\end{subequations}

From the above equations, the angles $\theta$ and $\overline{\theta}$ can be computed in terms of $\phi$, and the optimal value of $F_g$ is given by the optimal fidelity for tracking between pure qubit states, as described in Sec \ref{sec:ex31}. In particular, note that for the present problem, a valid indicator function is the one proposed in Eq. (\ref{eq:omegaequiv}),
\begin{equation}
\widetilde{\Omega}=2\arccos\left[\sin\left(2\phi\right)\right]-2\arccos\left[\sin^2\left(2\phi\right)\right]\leq 0\,,
\end{equation}
where the inequality holds for the specified range of $\phi$,
implying that the optimal fidelity is given by Eq. (\ref{eq:fidppunit}) with the proper values of $\theta$ and $\overline{\theta}$, explicitly
\begin{equation}
\mathcal{F}_{\rm HS}=\frac{1}{2}+\frac{1}{2}\sqrt{\pi_1^2+\pi_2^2+2\pi_1\pi_2\cos{\widetilde{\Omega}}}\,.
\end{equation}

For $\pi_1=\pi_2=1/2$, the above formula can be shown to be precisely the same as Eq. (38) of \cite{98Bruss2368}, which gives the optimal global fidelity of the cloner. Thus we have not only reproduced that previous result, but also determined how it is optimally modified to incorporate an unequal probability of preparation of $\ket{a}$ and $\ket{b}$.

Finally, let us just mention that the resulting optimal tracking unitary operation (call it $W$) is not quite the optimal cloning unitary operation $U$ appearing in Eq. (\ref{eq:Fg}) and detailed in \cite{98Bruss2368} ($U$ and $W$ do not even act in Hilbert spaces of equal dimensions).
Instead, $W$ constrains how $U$ acts on the states of the form $\ket{\psi 0}$, but to fully specify $U$ we would need to choose $U\ket{11}$ and $U\ket{01}$ such that $U$ is a unitary matrix. Since this choice is not unique and does not affect the fidelity, we can say that $W$ contains all the essential information associated with the optimal cloning map.

\section{Tracking with a Control Loop}\label{sec:quantctrl}

Although the strategy introduced in Section \ref{sec:strategy} has been tailored to correspond to a CPTP map, so far no insight on how such a map can be physically implemented has been given. In this section we provide a realization in terms of a quantum control scheme. Namely, procedures A and B are shown to have the structure of closed and open loop control, respectively.\par

We start by giving a possible Kraus decompositions for the CPTP maps representing our strategy. This is relevant here because the Kraus form of a CPTP map enables us to interpret that map as some generalized quantum measurement (with no record of the outcomes) \cite{00nielsen}. For $\Omega > 0$, the transformation $\mathcal{C}(\rho_i)=U \mathcal{D}(V \rho_i V^\dagger)U^\dagger$ from procedure A can be written as
\begin{equation}
\mathcal{C}(\rho_i)=(U M_1 V)\rho_i(U M_1 V)^\dagger+(U Y M_2 V)\rho_i(U Y M_2 V)^\dagger\,,
\end{equation}
with
\begin{subequations}\label{eq:meas_op}
\begin{align}
M_1&=\cos{\left(\frac{\chi-\eta}{2}\right)}\ket{+}\!\bra{+}+\sin{\left(\frac{\chi+\eta}{2}\right)}\ket{-}\!\bra{-}\,,\\
M_2&=\sin{\left(\frac{\chi-\eta}{2}\right)}\ket{+}\!\bra{+}-\cos{\left(\frac{\chi+\eta}{2}\right)}\ket{-}\!\bra{-}\,,
\end{align}
\end{subequations}
where  $\chi$ and $\eta$ are defined such that $\sin{\chi}=\mu_3$, $\cos{\chi}=\sqrt{1-\mu_3^2}$, $\sin{\eta}=\mu_2$ and $\cos{\eta}=\sqrt{1-\mu_2^2}$.

 For $\Omega \leq 0$, the transformation $\mathcal{C}(\rho_i)=U V \rho_i V^\dagger U^\dagger$ from procedure B is automatically in Kraus form, with a single Kraus operator $U V$.\par
We interpret these results as follows. First for $\Omega > 0$, the unitary $V$ is applied to the system and then a generalized quantum measurement with operators $M_1$ and $M_2$ is performed. Conditioned on observing the outcome `2', a Pauli $Y$ is applied to the system, followed by the unitary $U$. If the outcome is `1', the unitary $U$ is applied straight away. Due to this measurement-dependent dynamics (feedback), procedure A can be regarded as a closed loop control scheme.\par
\begin{figure}[h]
\includegraphics[width=8cm]{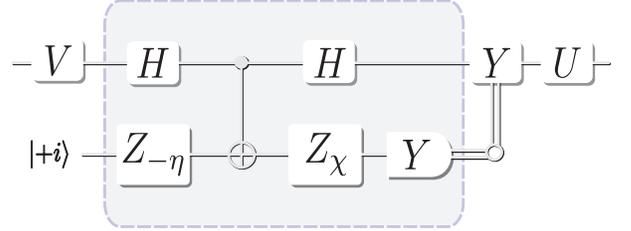}

\caption{A circuit model illustrating the feedback structure of procedure A. In the figure, $\ket{\pm i}=(\ket{0}\pm i\ket{1})/\sqrt{2}$ are the eigenvectors of the Pauli matrix $Y$, $H$ is the Hadamard gate and $Z_\theta=\exp{(-i\theta Z/2)}$. The highlighted circuit entangles the main system with the ancilla and projectively measures the ancilla in the basis $\{\ket{+i},\ket{-i}\}$. This induces a non-projective dynamics of the main system, and for this reason this block is referred to as a ``weak measurement''. If the measurement outcome is `$+i$', then the unitary transformations $Y$ and $U$ are applied to the main system; otherwise, only $U$ is applied.}\label{fig:circuit}
\end{figure}

Note that the measurement operators $M_1$ and $M_2$ are not projections, so the implementation of such a measurement requires the enlargement of the Hilbert space (by interaction with an ancilla), with subsequent (projective) measurement of the ancilla. Fig. \ref{fig:circuit} shows a possible circuit model for procedure A.

For $\Omega \leq 0$, there is clearly no measurement involved, hence the control strategy is implemented independent
of acquiring extra information from the system. For this reason, procedure B can be regarded as an open loop control scheme.

\section{Discussion and Conclusions}\label{sec:discussion}

In this paper we have introduced a simple quantum version of
a common classical control problem named tracking. Our quantum tracking problem consists of determining how to optimally enforce a certain dynamics to a qubit system, when the initial preparation of the qubit is uncertain (as modeled by a pair of states occurring with given prior probabilities) and the desired dynamics depends on the actual preparation. We presented an optimal quantum tracking strategy.\par

The tracking problem studied here is sufficiently general to provide an unifying approach to many problems in quantum information science as special cases. For example, some cases of quantum state discrimination, quantum state purification, stabilization of qubits against noise and state-dependent quantum cloning were explicitly shown to be instances of quantum tracking. As such, previously known quantum limits in the realization of these tasks were recovered via the application of our tracking strategy. Likewise, our tracking strategy can be used to obtain new and improved limits in the realization of other impossible quantum machines.

The derivation of our strategy was largely dependent on the fact that our figure-of-merit (the averaged Hilbert-Schmidt inner product) is linear in $\mathcal{C}$, which, in turn, is constrained to be an element of the \emph{convex} set of CPTP maps acting on qubits. This implies that the optimal map $\mathcal{C}$ belongs to the subset of extreme points, which has been fully characterized in \cite{02Ruskai159}. Thanks to a parametrization of these extreme points, the resulting optimization problem could be handled analytically when a few mild assumptions (supported by numerical observation) were made about the form of the optimal solution. The optimality was safeguarded \emph{a posteriori} via an argument based on the SDP structure of the tracking problem.\par

Analytical solutions for generalizations of the tracking problem studied here (e.g., other figures-of-merit and/or larger dimensional quantum systems) seem to require a modified approach from the one adopted here. For example, had we chosen to proceed with a better motivated figure-of-merit for mixed targets, such as the average fidelity, we would still have the guarantee that the optimal $\mathcal{C}$ is an extreme point, however optimality results about a possible guess would be harder to derive, since it is not known if/how the resulting optimization problem can be cast as a SDP when source and target states are mixed. Alternatively, we could have chosen, for example, to minimize the average trace distance, which can be cast as a SDP \cite{01Fazel,07Recht}. However, the trace distance is convex in $\mathcal{C}$, in which case its minimum is not an extreme point. Finally, had we kept our linear figure-of-merit but generalized from qubits to qudits for $d>2$ (or to multiple qubits), we would face the problem that the extreme points of the set of CPTP maps on higher dimensional matrix algebras are not well characterized.

A possibly simpler generalization is to preserve low dimensionality of the quantum system and linearity in the figure-of-merit, but allow for a larger number of possible sources and targets. In principle, this problem can be approached following exactly the same lines as adopted here. In fact, it is not difficult to see that a particular case of this more general problem can already be considered solved given the results of this paper. Consider we are given two sets $\mathcal{S}_1$ and $\mathcal{S}_2$, respectively with $n_1$ and $n_2$ elements (let $N = n_1 + n_2$), of qubit density matrices $\tau_j$ ($j=1,\ldots,N$), and want to send every element of $\mathcal{S}_i$ to $\overline{\rho}_i$ for $i=1,2$. In analogy with Eq. (\ref{eq:J}), define the figure-of-merit
\begin{equation}\label{eq:figmeritger}
\mathcal{F}_{\rm HS}=\sum_{j=1}^{n_1}{q_j{\rm Tr}\left[\mathcal{C}(\tau_j),\overline{\rho}_1\right]}+\sum_{j=n_1+1}^N{q_j{\rm Tr}\left[\mathcal{C}(\tau_j),\overline{\rho}_2\right]}\,,
\end{equation}
where the positive numbers $q_j$ set the priorities of each transformation, and $\sum_{j=1}^N q_j=1$. Due to the linearity of the trace and of quantum operations, Eq. (\ref{eq:figmeritger}) can be rewritten exactly as Eq. (\ref{eq:J}) with $\pi_1 = \sum_{j=1}^{n_1}{q_j}$, $\pi_2 = \sum_{j=n_1+1}^{N}{q_j}$,
\begin{align}
\rho_1=\frac{1}{\pi_1}\sum_{j=1}^{n_1}{q_j\tau_j} &\quad \mbox{and} \quad \rho_2=\frac{1}{\pi_2}\sum_{j=n_1+1}^{N}{q_j\tau_j}\,.
\end{align}
Note that $\pi_1,\pi_2\geq 0 $, $\pi_1+\pi_2=1$ and $\rho_1$, $\rho_2$ are valid density matrices. So, for $i=1,2$ and $j=1,\ldots,N$, the problem of optimally approximating the $N$-state transformation $\mathcal{S}_i \to \overline{\rho}_i$ with priority $q_j$ is equivalent to optimally approximating the $2$-state transformation $\rho_i\to\overline{\rho}_i$ with priority $\pi_i$.\par

\begin{acknowledgments}
  We acknowledge an anonymous referee for useful suggestions that helped to improve the original manuscript. P.E.M.F.M. thanks Yeong-Cherng Liang and C. J. Foster for helpful discussions and the support of the Brazilian agency Coordena\c c\~ao de Aperfei\c coamento de Pessoal de N\' ivel Superior (CAPES). This project was supported by the Australian Research Council.
\end{acknowledgments}

\appendix

\section{Perfect Tracking Conditions}\label{app:ptc}
A theorem closely related to the aims of this paper has been proved by Alberti and Uhlmann \cite{80Alberti163}, consisting of a mathematical criterion for the existence of physical operations perfectly transforming between pairs of qubit states. In this appendix we briefly review this theorem and prove an important corollary that is used in a number of places in this paper (e.g., sections \ref{sec:indicator} and \ref{sec:ex3}).
\begin{theorem}[Alberti and Uhlmann]\label{teo1}
Let $\rho_1$, $\rho_2$, $\overline{\rho}_1$, $\overline{\rho}_2$ be $2 \times 2$ density matrices. Then there exists a CPTP map $\mathcal{A}$ such that
\begin{equation}
\overline{\rho}_1=\mathcal{A}(\rho_1)\qquad\mbox{and}\qquad\overline{\rho}_2=\mathcal{A}(\rho_2)\,,
\end{equation}
if and only if
\begin{equation}\label{eq:alberti}
\|\overline{\rho}_1-t\overline{\rho}_2\|_{\rm tr}\leq\|\rho_1-t\rho_2\|_{\rm tr}\quad\mbox{for all }t\in \mathbb{R}^+\,.
\end{equation}
\end{theorem}
\noindent where $\|\cdot\|_{\rm tr}$ denotes the trace norm. As pointed out by Chefles, Jozsa and Winter \cite{04chefles11},
the condition (\ref{eq:alberti}) is equivalent
to the requirement that the target states are no more distinguishable than the source states by minimum error probability discrimination (Helstrom \cite{76Helstrom}), for any prior probabilities. In the particular case where $\overline{\rho}_1$ and $\overline{\rho}_2$ are pure states, this just means that the Bloch angle between $\overline{\rho}_1$ and $\overline{\rho}_2$ is smaller than the angle between $\rho_1$ and $\rho_2$. This is proved in the following.
\begin{corollary}\label{cor2}
Let $\overline{\rho}_1$ and $\overline{\rho}_2$ be any two pure distinct qubit states separated by an angle $\overline{\Theta}\in(0,\pi]$ in the Bloch representation. Let $\rho_1$ and $\rho_2$ be any (mixed or pure) qubit states separated by $\Theta\in(0,\pi]$. A CPTP map $\mathcal{A}$ such that
\begin{equation}
\overline{\rho}_1=\mathcal{A}(\rho_1)\qquad\mbox{and}\qquad\overline{\rho}_2=\mathcal{A}(\rho_2)
\end{equation}
exists if and only if $\rho_1$ and $\rho_2$ are also pure and $\overline{\Theta}\leq\Theta$.
\end{corollary}
\begin{proof}
First note that the inequality (\ref{eq:alberti}) can be equivalently written with both sides
squared. Also, since $\overline{\rho}_1-t\overline{\rho}_2$ and $\rho_1-t\rho_2$ are hermitian matrices, their trace
norm can be computed as the sum of their eigenvalues. In terms of the Bloch parameters, a straightforward computation gives
\begin{equation}
\|\overline{\rho}_1-t\overline{\rho}_2\|^2=4(1+t^2-2t\cos{\overline{\Theta}})\,, \label{eq:lhs}\\
\end{equation}
where we have made use of the fact that $t\in\mathbb{R}^+$, and
\begin{multline}
\|\rho_1-t\rho_2\|^2=\\
2\left[(1-t)^2+(R_1^2+t^2R_2^2-2 t R_1 R_2\cos{\Theta})\right]\\
+2\left|(1-t)^2-(R_1^2+t^2 R_2^2-2 t R_1 R_2\cos{\Theta})\right|\,, \label{eq:rhs}
\end{multline}
where $R_i$ gives the magnitude of the Bloch vector for $\rho_i$, $i=1,2$.\par

Now assume that the absolute value on the right hand side of Eq. (\ref{eq:rhs}) can be removed, then the
inequality (\ref{eq:alberti}) takes the form
\begin{equation}\label{eq:ineq_impossible}
1+t^2-2t\cos{\overline{\Theta}}\leq(1-t)^2\,,
\end{equation}
which for all $t\in\mathbb{R^+}$ is satisfied if and only if $\cos{\overline{\Theta}}=1$. However, as the (pure) target states are required to be distinct, we must have $\cos{\overline{\Theta}}<1$. As a result, the inequality (\ref{eq:ineq_impossible}) is never satisfied.\par

Assume then the complementary case (when the absolute value of Eq. (\ref{eq:rhs}) is removed
at the cost of a change of sign). Then (\ref{eq:alberti}) can be written as $F(t)\leq 0$ with
\begin{equation}
F(t)=(1-R_2^2)t^2-2t(\cos{\overline{\Theta}}-R_1 R_2\cos{\Theta})+(1-R_1^2)\,.
\end{equation}
If $R_2 \neq 1$, $F(t)$ is a strictly convex function of $t$, therefore cannot be bounded from above by $0$ for all $t\in\mathbb{R^+}$, so it is necessary that $R_2=1$ ($\rho_2$ must be pure). Then, define $G(t)=\left.F(t)\right|_{R_2=1}$, explicitly
\begin{equation}
G(t)=-2t(\cos{\overline{\Theta}}-R_1\cos{\Theta})+(1-R_1^2)\,,
\end{equation}
and require $G(t)\leq 0$.\par
If $R_1\neq 1$, $G(t)$ is a linear function of $t$ with strictly positive linear coefficient. Again, such a function cannot be bounded from above by $0$ for all $t\in\mathbb{R^+}$, so it is necessary to make $R_1=1$ ($\rho_1$ must be pure). Finally, define $H(t)=\left.G(t)\right|_{R_1=1}$, i.e.,
\begin{equation}
H(t)=-2t(\cos{\overline{\Theta}}-\cos{\Theta})\,,
\end{equation}
and require $H(t)\leq 0$. Clearly, this inequality is satisfied for all $t\in\mathbb{R^+}$ if and only if $\cos{\overline{\Theta}}\geq\cos{\Theta}$, or equivalently, $\overline{\Theta}\leq\Theta$.
\end{proof}
\section{Technical details}
\subsection{Properties of $S$ and $T$}\label{app:ST}
Here, we prove that $S>0$ and $S+T>0$ if and only if one of the following holds
 \begin{enumerate}
\item[i)] $\{\overline{\bm{R}}_1,\overline{\bm{R}}_2\}$ is linearly independent; or
\item[ii)] $\{\overline{\bm{R}}_1,\overline{\bm{R}}_2\}$ is linearly dependent with $T>0$.
 \end{enumerate}
 Moreover, we show that the complementary case
\begin{enumerate}
\item[iii)] $\{\overline{\bm{R}}_1,\overline{\bm{R}}_2\}$ is linearly dependent with $T\leq 0$,
\end{enumerate}
occurs only if $\Omega=0$.\par
 This result is useful to demonstrate that the coefficients $\mu_1$, $\mu_2$, $\mu_3$ and $s_1$ defined in Eq. (\ref{eq:nontrivial}) for $\Omega>0$ (procedure A) are always (a) well-defined, (b) real (c) within the range $[0,1]$. We start with the following lemma

\begin{lemma}\label{lemma1}
Let $\bm{R}_1$, $\bm{R}_2$ be real three dimensional vectors such that $R_i\leq 1$ $(i=1,2)$. Define $\bm{R}_-\mathrel{\mathop:}=\bm{R}_1-\bm{R}_2$. If $R_-\neq0$ (i.e., $\bm{R}_1$, $\bm{R}_2$ are distinct), then $R_-^2>R_\times^2$.
\end{lemma}
\begin{proof}
Consider the triangle defined by the vectors $\bm{R}_1$, $\bm{R}_2$ and $\bm{R}_-$ as shown in Fig. \ref{fig:triangle}.
\begin{figure}
\includegraphics[width=5cm]{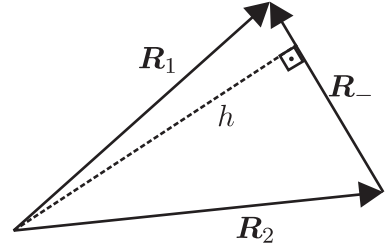}
\caption{Schematic for proof that $R_- > R_\times$}\label{fig:triangle}
\end{figure}
The magnitude of $\bm{R}_\times$ gives twice the area of the triangle so that
\begin{equation}
R_\times^2=h^2R_-^2\,,
\end{equation}
where $h$ is the altitude relative to the side of length $R_-$. We write the following
\begin{equation}
R_\times^2=h^2R_-^2\leq \min{\left(R_1,R_2\right)}^2 R_-^2\leq R_-^2\,.\label{eq:ineqs}
\end{equation}

The first inequality is a direct consequence of the Pythagorean theorem, and the second follows from $R_i\leq 1$. This establishes that $R_-^2 \geq R_\times^2$. This inequality is trivially saturated if $R_-=0$. To see that this is the only case where saturation occurs, assume $R_-\neq 0$ and require saturation of both inequalities in Eq. (\ref{eq:ineqs}). The first inequality is saturated iff $R_-^2=|R_1^2-R_2^2|$ (by the Pythagorean theorem), and the second one iff $R_1=R_2=1$. Taken together, these conditions imply $R_-=0$, which contradicts the hypothesis. Therefore, if $R_-\neq 0$ (i.e., $\bm{R}_1\neq\bm{R}_2$), then  $R_-^2 > R_\times^2$.
\end{proof}
Now, recall that
\begin{align}
S = \sqrt{T^2+4\overline{R}_\times^2(R_-^2-R_\times^2)}\,.
\end{align}
Assume first linear independence of $\{\overline{\bm{R}}_1,\overline{\bm{R}}_2\}$ (i.e., $\overline{R}_\times\neq 0$). From Lemma \ref{lemma1}, it is immediate that $S>0$. Moreover,
\begin{align}
S+T = T+\sqrt{T^2+4\overline{R}_\times^2(R_-^2-R_\times^2)}>T+|T|\geq 0\,,
\end{align}
where the first inequality follows from Lemma \ref{lemma1} and the second is trivial. Therefore, $S>0$ and $S+T>0$ if condition (i) holds.\par

 For linearly dependent $\{\overline{\bm{R}}_1,\overline{\bm{R}}_2\}$, it is easy to see that $S=|T|$ and $S+T=|T|+T$, therefore $S>0$ and $S+T>0$ if condition (ii) holds.\par

  To prove the \emph{only if} part, consider the complementary case (iii). It is immediate that $S+T=0$ if $\{\overline{\bm{R}}_1,\overline{\bm{R}}_2\}$ are linearly dependent and $T \leq 0$, hence (i) and (ii) are the only situations where the premise holds.\par
  It follows trivially from the discussion above that $\Omega=0$ for condition (iii). Simply note that $S+T=0$ and the linear dependence of the targets Bloch vectors requires $2R_\times \overline{R}_\times=0$.

\subsection{Well-definedness of the dual feasible point}\label{app:welldef}

The proposed values for the coefficients $\mathfrak{x}_1$ and $\mathfrak{x}_3$ defined in Eqs. (\ref{eq:xi3p}), (\ref{eq:xi1m}) and (\ref{eq:xi3m}) have the quantities  $\Gamma_a$ and $\Gamma_b$ appearing in the denominator. In this appendix we show that this does not lead to any singularity as long as the indicated range of $\Omega$ is considered.

 To see that, note that $\Gamma_a=0$ if and only if $\overline{R}_+=\overline{R}_\times=0$. This, in turn, is equivalent to the statement that the targets have Bloch vectors of same magnitude $\overline{R}$ pointing to opposite directions, which used in Eq. (\ref{eq:T}) gives $T=-R_-^2\overline{R}^2$. In these circumstances, $\Omega$ can be easily computed to be $\Omega=T+|T|=0$. Therefore, no singularity can occur in Eq. (\ref{eq:xi3p}) in the range $\Omega>0$.\par

 Similarly, $\Gamma_b=0$ if and only if $\overline{R}_+^2=T-2R_\times\overline{R}_\times$, in which case we can write $\Omega=S+\overline{R}_+^2$. In the sequence we show that $S+\overline{R}_+^2>0$, thus no singularity can occur in Eqs. (\ref{eq:xi1m}) and (\ref{eq:xi3m}) in the range $\Omega\leq 0$.\par

  From the definition of $S$ in Eq. (\ref{eq:S}), it is immediate that the inequality $S+\overline{R}_+^2\geq 0$ holds, so we just need to show that $S+\overline{R}_+^2\neq 0$. Suppose, on the contrary, that $S=-\overline{R}_+^2$, which is possible only if $S=\overline{R}_+=0$. From Eq. (\ref{eq:S}), this can be seen to be equivalent to $T=\overline{R}_\times=\overline{R}_+=0$. To see that this leads to a contradiction, use once again the fact that $\overline{R}_\times=\overline{R}_+=0$ implies opposing target Bloch vectors of same magnitude $\overline{R}$, which gives $T=-R_-^2\overline{R}^2\neq 0$. The inequality follows from the conditions of the problem: the source states cannot be identical ($R_-\neq 0$), and the case where the two targets are identical to the maximally mixed states has been excluded from the analysis ($\overline{R}\neq 0$).

\subsection{Characteristic Polynomials for $F$}\label{app:charpoly}

In this appendix we compute the characteristic polynomials of the matrix $F$, Eq. (\ref{eq:lmi}), with the set of coefficients given in Eq. (\ref{eq:coefomegap}) (for $\Omega > 0$, procedure A) and Eq. (\ref{eq:coefomegam}) (for $\Omega \leq 0$, procedure B). By studying these polynomials, we show that $F\geq 0$, thus completing the proof of the optimality of our tracking strategy.

\subsubsection{Procedure A}
For the set of coefficients (\ref{eq:coefomegap}) (case $\Omega>0$), the characteristic equation for $F$ factors as
$\lambda^2 P_2(\lambda)=0$, where
\begin{equation}
P_2(\lambda)=\lambda^2-\Gamma_a\lambda+\upsilon\left[(R_-^2-R_\times^2)\Gamma_a^4-\Xi^2\right]\,,\label{eq:nont_poly}
\end{equation}
and
\begin{equation}
\upsilon=\frac{4 R_-^2 \overline{R}_\times^2+(S+T)^2}{8 R_-^2 \Gamma_a^2 S(S+T)}\,.
\end{equation}
Since both $\Gamma_a$ and $\upsilon$ are positive, the eigenvalues of $F$ are non-negative if the term in square brackets in the Eq. (\ref{eq:nont_poly}) is non-negative when $\Omega > 0$. We now show that this term is non-negative irrespective of the sign of $\Omega$.\par

First use Eq. (\ref{eq:xixi}) to substitute for $\Xi^2$, after some manipulation we find that
\begin{multline}\label{eq:computstep1}
(R_-^2-R_\times^2)\Gamma_a^4-\Xi^2=\\
R_-^2\left[a\left(\overline{R}_+^2+\frac{R_-^2\overline{R}_\times^2}{S+T}\right)+ R_-^2\overline{R}_\times^2+\overline{R}_+^2T\right]\,,
\end{multline}
where we have defined
\begin{equation}
a\mathrel{\mathop:}=\frac{4\overline{R}_\times^2}{S+T} \left(R_-^2-R_\times^2\right)=S-T\,,
\end{equation}
with the second equality following from Eq. (\ref{eq:S}). Note that the non-negativity of $(R_-^2-R_\times^2)\Gamma_a^4-\Xi^2$ cannot be immediately concluded from Eq. (\ref{eq:computstep1}) --- although the first and second summands are non-negative, the term $\overline{R}_+^2 T$ does admit negative values. However, using $a=S-T$ in Eq. (\ref{eq:computstep1}), after some rearrangement we get,
\begin{multline}
(R_-^2-R_\times^2)\Gamma_a^4-\Xi^2=\\
R_-^2\left[S\overline{R}_+^2+R_-^2\overline{R}_\times^2\left(1+\frac{S-T}{S+T}\right)\right]\geq 0\,,
\end{multline}
from which the fulfillment of the inequality is obvious. In conclusion, procedure A is optimal.\par

\subsubsection{Procedure B}
For the set of coefficients (\ref{eq:coefomegam}) (case $\Omega\leq 0$), the characteristic equation for $F$ is $\lambda P_3(\lambda)=0$, where
\begin{equation}
P_3(\lambda)=\lambda^3-\Gamma_b\lambda^2+\varpi\lambda+\omega\,,\label{eq:t_poly}
\end{equation}
and
\begin{align}
\varpi&=\tfrac{1}{4 R_-^4 \Gamma_b^2}\left[\left(1+\tfrac{R_\times \xi}{R_-^2\Gamma_b^2}\right)R_-^4\Gamma_b^4-(R_-^2+R_\times^2)(\xi^2+\Xi^2)\right],\label{eq:varpi}\\
\omega&=-\frac{(R_\times\Gamma_b^2-\xi)\left[R_-^2\Gamma_b^2\xi-R_\times(\xi^2+\Xi^2)\right]}{8 R_-^4\Gamma_b^3}\label{eq:omega}\,.
\end{align}
from which it follows that the eigenvalues of $F$ are non negative if $\varpi \geq 0$ \emph{and} $\omega \leq 0$ \emph{when} $\Omega \leq 0$. Next, we simplify Eqs. (\ref{eq:varpi}) and (\ref{eq:omega}) in order to make it clear that these conditions are satisfied.\par

It is just a matter of applying Eqs. (\ref{eq:xi}) and (\ref{eq:xixi}) to Eq. (\ref{eq:varpi}) to show that
\begin{equation}
\varpi=\tfrac{1}{4}\left(-\Omega+S+R_\times\overline{R}_\times\right)\geq 0\,,
\end{equation}
from which the inequality is clearly seen to hold if $\Omega \leq 0$.\par

To prove that $\omega\leq 0$ if $\Omega \leq 0$, consider first the term in the square brackets in Eq. (\ref{eq:omega}). Again, employing Eqs. (\ref{eq:xi}) and (\ref{eq:xixi}) this can be simplified to   $R_-^4\overline{R}_\times\Gamma_b^2$ which is obviously non-negative. Therefore, the validity of the inequality $\omega \leq 0$ if $\Omega\leq 0$ is now solely conditioned on the validity of the inequality
\begin{equation}\label{eq:ineqprocB}
R_\times\Gamma_b^2-\xi \geq 0\quad\mbox{for}\quad \Omega \leq 0\,.
\end{equation}
To see that this is so, first note that the only way to satisfy the conditions $\overline{R}_\times=0$ and $\Omega\leq 0$ is to have $S=T=\Omega=0$, which implies that (\ref{eq:ineqprocB}) is satisfied with saturation. Consider then the complementary case $\overline{R}_\times\neq 0$ and $\Omega \leq 0$. Using Eq. (\ref{eq:xi}) for $\xi$, and multiplying and dividing by $4\overline{R}_\times$, we get
\begin{equation}\label{eq:computstep2}
R_\times\Gamma_b^2-\xi=\frac{1}{4\overline{R}_\times}\left[-4 R_\times\overline{R}_\times(\Omega-S)-4\overline{R}_\times^2R_-^2\right]\,.
\end{equation}

Now, from Eq. (\ref{eq:S}), we know that $4\overline{R}_\times^2R_-^2=S^2-T^2+4R_\times^2\overline{R}_\times^2$, which used in Eq. (\ref{eq:computstep2}) gives, after some algebra,
 \begin{equation}
R_\times\Gamma_b^2-\xi=-\frac{\Omega}{4\overline{R}_\times}\left(S-T+2R_\times\overline{R}_\times\right)\geq 0\,.
 \end{equation}
Once again, the inequality is obviously true if $\Omega \leq 0$, thus establishing the optimality of procedure B.


\end{document}